\documentclass[twocolumn,showpacs,preprintnumbers,amsmath,amssymb,aps]{revtex4}
\usepackage{graphicx}
\begin{document}

\title{Dynamically Induced Locking and Unlocking Transitions in Driven Layered Systems with Quenched Disorder }  
\author{C. Reichhardt and C.J. Olson Reichhardt} 
\affiliation{ 
Theoretical Division, 
Los Alamos National Laboratory, Los Alamos, New Mexico 87545 USA}

\date{\today}
\begin{abstract}
Using numerical simulations, we examine a simple model of two or more 
coupled one-dimensional channels of driven particles with repulsive interactions
in the presence of quenched disorder.
We find that this model exhibits a remarkably rich variety of dynamical
behavior as a function of the strength of the quenched disorder, coupling between channels,
and external drive. 
For weaker disorder, the channels depin in a single step.  
For two channels we find dynamically
induced decoupling transitions 
that result in coexisting pinned and moving phases as well as
moving decoupled phases 
where particles in both channels move at different average
velocities and slide past one another. 
Decoupling can also be induced by changing the relative strength of the
disorder in neighboring channels.
At higher drives, we observe a dynamical recoupling or locking transition
into a state with no relative motion between the channels.
This recoupling produces unusual velocity-force signatures, including
negative differential conductivity.
The depinning threshold shows distinct changes near the decoupling and coupling
transitions and exhibits a peak effect
phenomenon of the type that has been 
associated with transitions from elastic to plastic flow in other systems. 
We map several dynamic phase diagrams showing the coupling-decoupling 
transitions and the regions in which hysteresis occurs. 
We also examine the coexistence regime for channels with unequal amounts of
quenched disorder.
For multiple channels, multiple coupling and decoupling transitions 
can occur; however, many of the general features found for the two channel 
system are still present.
Our results should be relevant to depinning in layered geometries 
in systems such as vortices in layered
or nanostructured superconductors and Wigner or 
colloidal particles confined in nano-channels; they are also
relevant to the general understating of plastic flow.  
\end{abstract}
\maketitle

\vskip2pc

\section{Introduction}
A collection of interacting particles on an ordered or disordered
substrate undergoes a depinning transition under an applied drive from
a pinned state with immobile particles to an elastic or plastic sliding state
\cite{Fisher,Fisher2,Sethna,Jensen,Higgins,Elder,Hu,Yang,CR,Ling}. 
In an elastic sliding state, each particle keeps the same neighbors over
time \cite{Fisher,Fisher2,Sethna}, while in a plastic sliding state,
the particles do not keep the same neighbors 
\cite{Jensen,Higgins,Ling,Kes,Watson,Koshelev,Moon}. 
Many systems show a transition from
elastic depinning for weak substrates 
to plastic depinning for strong substrates that is
associated with changes in the velocity force curves and 
the depinning threshold as well as the onset of strong
hysteretic effects \cite{Higgins,Kes,Olson,Menon}.    
For example, the peak effect phenomenon for the transport of superconducting
vortices, where a peak in the critical depinning force or critical current
appears as a function of temperature or magnetic field,
has been associated with a transition from elastic to plastic depinning and
is known to be related to the disordering of the vortex lattice 
\cite{Higgins,Kes}. 
Although the peak effect is primarily associated with superconductors,
it can in principle appear in other systems that undergo a
transition from elastic to plastic depinning.
Several recent studies of colloidal systems and frictional
systems have revealed peak effect-like behavior \cite{CR,Olson,Menon}.

Plastic depinning transitions exhibit a wide variety of characteristics
depending on whether the substrate is random
\cite{Jensen} or 
periodic \cite{CR,Elder,Xiao,Grier,Lacasta,Reichhardt,Xiao}. 
The resulting plastic flow
ranges from motion in winding channels to
avalanche behaviors with strong fluctuations, 
coexistence of large pinned and flowing regions,
or transitions from two-dimensional (2D)
mixing phases to one-dimensional (1D) decoupled channel phases. 
For higher drives, many systems undergo a transition from plastic flow to 
a more ordered flow state, 
such as a moving smectic state where the particles organize
into 1D channels that can slide past one another 
\cite{Balents,Moon,Olson} or partially ordered
moving crystal states \cite{Koshelev,Balents,Elder,Hu}.   

Due to the complexity of plastic depinning phenomena in 2D and 
three-dimensional (3D) systems, numerous 
simpler systems have been proposed that still 
exhibit plastic depinning, such as particles flowing in
a series of simple coupled layers. 
Many systems with depinning transitions 
can be modeled effectively as layered systems, including sliding
charge density waves \cite{Fisher,Nattermann}
and vortices in strongly layered superconductors \cite{Zimanyi,OlsonN}. 
Advances in fabrication techniques have made it possible to create 
nanostructured systems in which particles such as vortices move through
effectively 1D
coupled channels \cite{Drose}.
Marchetti {\it et al.} considered a mean field 
anisotropic slip model of coupled channels of particles 
oriented parallel to a driving force \cite{MC}.  Along each channel, the 
particle interactions are elastic, but slip can occur between
neighboring channels.
In this model, when the depinning is elastic the channels show
no slip or hysteresis, but for plastic depinning there is both
slip and hysteresis.
Further theoretical work on systems with
only two layers 
demonstrated that a transition from non-hysteretic elastic depinning to 
hysteretic plastic depinning occurs as a function of disorder \cite{Wiese}.
For 3D layered sliding charge density wave systems, 
a transition from an elastic depinning
to plastic hysteretic depinning has also been predicted \cite{Nattermann}, 
along with a second coupling transition at higher drives 
when the charge density waves begin to flow coherently. 
Numerical work on 3D layered superconducting vortex systems 
has shown a similar disorder-induced
transition from elastic to plastic flow 
marked by a large increase in the depinning
threshold when the layers decouple,
which is typical of a 
peak effect behavior \cite{Zimanyi,OlsonN,Zhao}. We note that 
although the transition to plastic 
depinning is marked by an increase in the 
depinning threshold, in real
superconductors the depinning threshold 
decreases as the superconducting critical temperature or 
field is approached due to the changing penetration
depth. This feature is generally not included in most simulations.      

In this work we consider a simple depinning model consisting of two or
more coupled 1D channels of repulsively interacting particles, 
all of which are subjected
to a driving force and quenched disorder.
This system could be experimentally realized 
using colloidal particles in coupled 1D channels with
a disordered substrate and driven by an external electric field 
\cite{Bechinger,Peeters,Farias,Goree},
coupled 1D wires containing Wigner crystal phases
\cite{Glasson,Rojo,Stopa,Piacente},
or vortices on corrugated substrates driven parallel
to the corrugation. 
Our simulations of this model include much more detail than can
be captured in mean field studies \cite{MC,Saunders,Wiese}. 
In addition to confirming many of the predictions from the theoretical studies,
we find a rich variety of new features.
We induce decoupling transitions by changing the relative strength of
the disorder from channel to channel, something that has not been considered
in previous work on the depinning of layered systems.
We also study commensuration effects by varying the particle density in
one channel relative to the density in other channels.
We drive all of the layers of particles; 
previous studies of this type of model considered transformer
geometries with the drive applied to only one layer \cite{C,CB}.
Despite the apparent simplicity of 
our model, we find that even the two layer system 
has a wide variety of dynamical phases and exhibits all the
salient features found for elastic and plastic depinning 
phenomena, including a peak effect at the transition between the 
two types of depinning.
We also show that the transport signatures of the dynamical coupling and
recoupling transitions can be enhanced in systems where the channels
do not all have the same amount of disorder,
suggesting that experimental realizations of this geometry may be
an excellent way to probe these dynamical states.   

The paper is organized as follows.  In Section II we describe our model 
and simulation technique.  Section III discusses decoupling transitions in
a two channel system as the spacing between channels is varied, as well as
a peak effect appearing at the transition from elastic to plastic
depinning.  In Section IV, we consider two channel samples in which the
strength of the disorder differs in the two channels, where we observe
drag-induced pinning as well as hysteretic velocity-force signatures.  
In Section V, we find commensuration effects when the number of particles 
in each channel is varied for a two channel system, and we also study
the effect of changing the particle-particle interaction prefactor.
We introduce an eight channel system in Section VI and show that as disorder
strength and channel coupling is varied, the behavior resembles that of
the two channel system.  We also study representative examples of
eight channel systems in which not all channels contain the same number
of particles.  We conclude with a summary in Section VII.

\section{Simulation}

\begin{figure}
\includegraphics[width=3.5in]{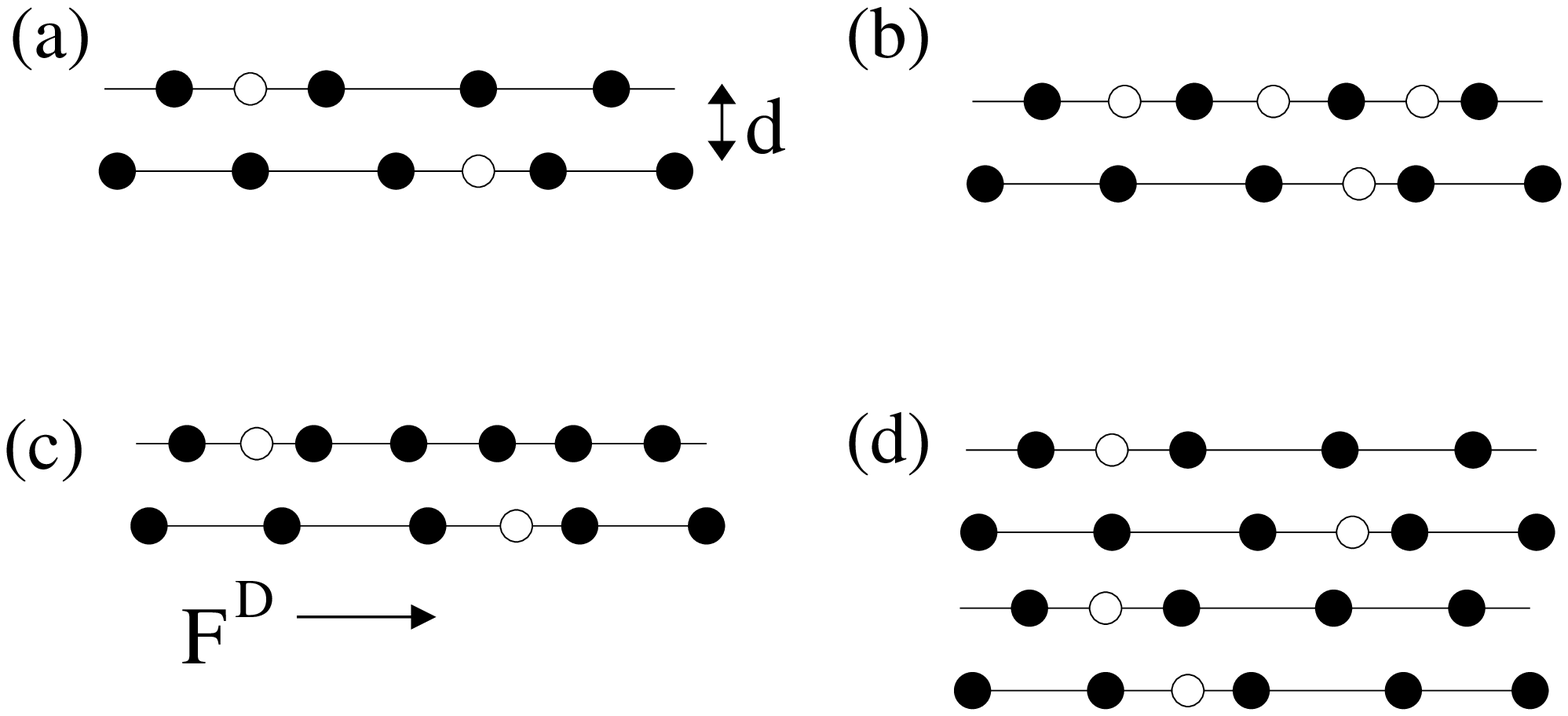}
\caption{
Schematic of the system. The particles (filled circles) are constrained to
move along 1D channels that are separated by a distance $d$.
Each channel contains 
$N_j$ particles as well as
$n_j$ randomly placed pinning sites (open circles) that
each have a maximum pinning force of $F^p_j$.
A uniform drive $F^D$ is applied to all particles in the positive
$x$ direction.
The particles interact via a repulsive Yukawa potential with other particles
in the same channels and in neighboring channels.
(a) A two channel system with $N_{1}=N_{2}$ and $n_{1}=n_{2}$.
(b) Two channels with an equal number of particles in each channel, 
$N_{1}=N_{2}$, but with more pinning in the upper channel, 
$n_{1}>n_{2}$. 
(c) Two channels with equal pinning in each channel, $n_{1}=n_{2}$, but more
particles in the upper channel, $N_{1}>N_{2}$. 
(d) Multiple channels each containing the same number
of particles, $N_j=N_1$, and pins, $n_j=n_1$.
\label{fig:imagefig}
}
\end{figure}

We model an array of $M$ coupled 1D channels separated by a
distance $d$ where channel $j$
contains $N_{j}$ particles \cite{C}. 
The total number of particles in the system is $N=\sum_{j=1}^M N_j$.
The particle-particle interactions are repulsive, and each particle
interacts both with particles in the same channel as well as with
particles in nearby channels.
The number of randomly placed pinning sites in channel $j$ is given
by $n_{j}$.
All of the particles in the system are subjected to a uniform external
driving force $F^D$.
We apply periodic boundary conditions in the $x$ direction, along the length
of the channels, and the total length of our system is $L=72a_0$, where
$a_0$ is our unit of length that is typically a micron in colloidal systems.
The boundaries are open along the $y$ direction transverse to the channels.
Fig.~\ref{fig:imagefig} shows a schematic
of our system in several different configurations.
In Fig.~\ref{fig:imagefig}(a) there are $M=2$ channels with equal
numbers of particles, $N_1=N_2$, and pins, $n_1=n_2$, in each channel.
Figure \ref{fig:imagefig}(b) shows a two channel system with equal numbers
of particles but with more pinning sites in the upper channel, $n_1>n_2$,
while in Fig.~\ref{fig:imagefig}(c) there are equal numbers of pins but
more particles in the upper channel, $N_1>N_2$.
Figure~\ref{fig:imagefig}(d) shows a multiple channel $M=5$ system with equal
numbers of particles in each channel, $N_j=N_1$, and equal numbers of pins in
each channel, $n_j=n_1$.

The particles interact via a Yukawa or screened Coulomb potential that
is appropriate for charged colloids or for 
charge transport in the presence of screening.
We show later that the same general dynamic phases also 
occur for repulsively interacting vortices in type-II superconductors.
For the colloidal system, the 
dynamics of a particle $i$ is determined by integrating the 
following overdamped equation
of motion:
\begin{equation}
\eta\frac{ d{\bf R}_{i}}{dt} = {\bf F}_{i}^{pp}  + {\bf F}^{s}_{i} + 
{\bf F}^{D} + {\bf F}^{T}_{i} \ .
\end{equation}
Here ${\bf R}_{i}$ is the location of particle $i$ and
$\eta$ is the damping  
coefficient that is set equal to one.  
The 
particle-particle interaction force is
${\bf F}^{pp}_{i} = -\sum_{j\ne i}^{N} \nabla V(R_{ij}){\hat {\bf R}}_{ij}$    
where 
$V(R_{ij}) = (q^2E_{0}/R_{ij})\exp(-\kappa R_{ij})$, 
$R_{ij}=|{\bf R}_i-{\bf R}_j|$,
${\hat {\bf R}}_{ij} = ({\bf R}_{i} - {\bf R}_{j})/R_{ij}$, 
$E_{0} = Z^{*2}/4\pi \epsilon\epsilon_{0}$,
$\epsilon$ is the solvent dielectric constant, 
$Z^{*}$ is the effective charge of a colloidal particle, and
$q^2$ is the dimensionless squared colloid charge that is taken to be
$q^2=2.0$ unless otherwise noted.
We take the screening length $1/\kappa=4a_0$ and
choose the distance between channels in the range
$d\leq 3a_0$ (equivalent to $d/\kappa \leq 2$)
to ensure that particles in neighboring channels interact with each other.
In the absence of pinning, the particles in an isolated channel would
adopt a lattice spacing of $a_j=N_j/L$.
The substrate pinning force is given by
${\bf F}^{s}_{i}=\sum^{n_j}_{k = 1}F^{p}_j(R^{p}_{ik}/R_{p})\Theta(R_{p} - R^{p}_{ik}){\hat {\bf R}_{ij}^{p}}$.  Here particle $i$ sits in
channel $j$, $n_j$ is the number of pinning sites in channel $j$,
$\Theta$ is the Heaviside step function, 
${\bf R}_{k}^{p}$ is the location of pinning site $k$, 
$R_{ik}^{p}  = |{\bf R}_{j} - {\bf R}_{k}^{p}|$,
and ${\bf \hat R}_{ik}^{p}=({\bf R}_j-{\bf R}_k^{p})/R^{p}_{ik}$. 
The pinning site radius is $R_{p}$
and the maximum strength of the
pinning sites in channel $j$ is
$F^{p}_j$.  Within each channel the pins are placed in randomly chosen
non-overlapping positions that are different for each channel.
This pinning model has been used in previous simulations
to represent quenched disorder
for colloidal systems, classical electron systems, 
and vortices in superconductors.   
The external drive ${\bf F}^{D}=F^D{\bf \hat x}$ 
is applied in the positive $x$-direction. The drive is increased
in increments of $\delta F^D = 0.001$, 
and after each increment $F^D$ is held 
fixed for $10^{5}$ simulation time steps to ensure that any transient
effects have subsided.
The thermal force ${\bf F}^{T}_i$ 
is modeled by adding a Langevin noise term with the following
properties: $\langle F^{T}_i(t)\rangle = 0$ 
and $\langle F^{T}_i(t)F^{T}_j(t^{\prime})\rangle = 2\eta k_{B}\delta_{ij}\delta(t-t^\prime)$, where
$k_{B}$ is the Boltzmann constant. 
We initially consider the $T=0$ case.
We measure the average particle velocity in each channel for every
force increment,
$V_{j} = \langle N_{j}^{-1}\sum_{i=1}^{N_{j}}{v_i}\rangle$,
where $v_i=(d{\bf R}_i/dt) \cdot {\bf \hat x}$.
We also measure the net velocity of all the channels,
$V=\sum_{i=1}^M V_i$.

\begin{figure}
\includegraphics[width=3.5in]{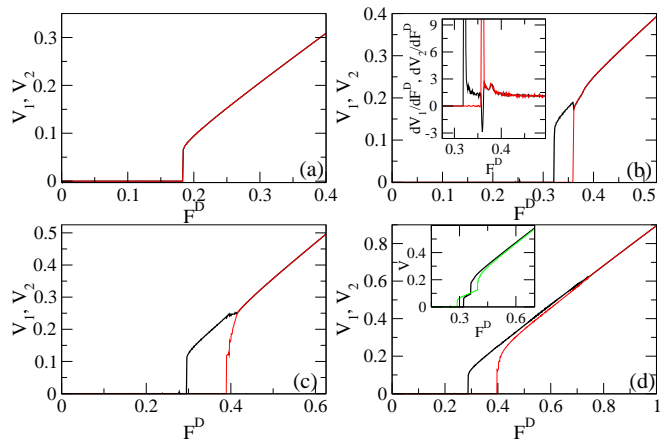}
\caption{
The velocity of each channel, $V_1$ (dark line)
and $V_2$ (light red line), in a two channel ($M=2$) system
versus applied drive $F^D$ with fixed quenched disorder
for varied distance $d$ between the channels. 
Each channel has the same number of particles, 
$N_1=N_2$, and pinning sites $n_1=n_2$.
Here $n_{1}/N_1 = 0.25$, 
$F^{p}_1 = F^{p}_2 = 6.0$,  $R_{p} = 0.23a_0$,
and the in-channel lattice constant $a_1=a_2=L/N_1=1.5a_0$.    
(a) For 
$d/a_1 = 1.133$, 
there is an elastic depinning transition and both
channels depin simultaneously.
(b) At 
$d/a_1 = 1.47$, 
channel 1 depins first while
channel 2 remains pinned. Once channel 2 depins, the
channels become dynamically locked. 
The locking transition is associated with negative differential conductivity
(NDC)
in $V_1$, as illustrated in the inset where
$dV_{1}/dF^{D} < 0.0$ at $F^D=0.36$. 
(c) For 
$d/a_1 = 1.57$ 
near $F^D=0.39$ there is a region where
both channels are depinned but $V_1 \neq V_2$,
indicating that the channels
are sliding past one another. 
Dynamical locking occurs for $F^{D} > 0.415$. 
(d) At 
$d/a_1 = 1.64$, 
the recoupling transition is shifted to much
higher $F^D$.
Inset: The sum of the velocities $V=V_1+V_2$ for $d/a_1=1.47$ (dark line)
and $d/a_1=1.64$ (light green line) shows the occurrence of two step
depinning; there is no NDC in $V$ for the $d/a_1=1.47$ sample.
}
\label{fig:vel2}
\end{figure}

\section{Decoupling Transitions for Two Layer Systems } 
We first consider a two layer system in which
we vary the interlayer distance $d$.
We can effectively decrease the coupling between the
layers by increasing $d$ 
since the interaction between Yukawa particles becomes weaker
for increasing interparticle distance. 
In Fig.~\ref{fig:vel2} we plot the velocity force curves for 
channels 1 and 2 in a system with 
equal numbers of particles in each channel ($N_1=N_2$), in-channel
lattice constant $a_1=a_2=1.5a_0$, equal numbers of pins in
each channel ($n_1=n_2$), 
$n_1/N_1 = 0.25$,
$F^{p}_1 = F^p_2 = 6.0$, and $R_{p} = 0.23$ 
for varied $d$ measured in terms of $d/a_1$. 
For small $d$ or strong coupling, the two 
channels depin simultaneously without any slipping between
the channels, as shown in Fig.~\ref{fig:vel2}(a) for $d/a_1=1.333$. 
All the particles keep the same neighbors, 
indicating that the depinning is elastic. 
The value of $F^D$ at which depinning first occurs is 
termed the critical depinning threshold, $F_c$.

As $d$ increases
there is a transition to a state where the channels depin individually, 
as shown in Fig.~\ref{fig:vel2}(b) for $d/a_1 = 1.47$. 
Here channel 1 depins at $F^{D} = 0.32$ while channel 2
remains pinned until $F^{D} = 0.36$. 
Since both channels have equal numbers of pins, equal numbers of particles,
and experience the same driving force, the difference in depinning threshold
arises due to the random placement of the pinning sites in the two channels,
which differs from channel to channel.
As soon as channel 2 depins,  the system transitions directly into
a state where the moving channels
are dynamically locked with each other and
move at the same velocity, $V_1=V_2$, without any slipping. 
As the inset in Fig.~\ref{fig:vel2}(b) shows,
the dynamic locking transition coincides with a drop in $V_{1}$ 
indicative of 
{\it negative} differential conductivity (NDC), 
where a  system of particles shows a decrease in the velocity
under increasing drive,  
$dV_{1}/dF^{D} < 0.0$.  
The NDC occurs only in channel $1$ at $F^D=0.36$ when $V_2$ jumps from
zero to a finite value,
which also corresponds to a peak in $dV_{2}/dF^{D}$.

Fig.~\ref{fig:vel2}(c) shows the velocity force curves for  $d/a_1 = 1.57$, where the
depinning threshold for channel 1 is again lower than that for channel 2.
In this case, after channel 2 depins the system does not immediately
enter the moving locked phase.  Instead, we find a region
$ 0.39 < F^{D} < 0.415$ with $V_1 \neq V_2$, $V_1>0$, and $V_2>0$.
Here both channels are moving but at different average velocities, 
indicating that the particles in channel 1
are sliding past the particles in channel 2.
The channels
become dynamically locked for $F^{D} > 0.415$, where $V_1=V_2$.
As $d/a_1$ is further increased, the dynamical locking transition
shifts to larger values of $F^D$.  For example, in 
Fig.~\ref{fig:vel2}(d) for $d/a_1 = 1.64$, 
the channels are only dynamically coupled for $F_{D} > 0.742$.  

In the inset of Fig.~\ref{fig:vel2}(d) we plot the overall velocity
$V=V_1+V_2$ versus $F^D$ for the samples with $d/a_1=1.47$ and $d/a_1=1.64$
from Fig.~\ref{fig:vel2}(b) and Fig.~\ref{fig:vel2}(d), respectively.
These are the curves that would be obtained in an experimental measurement
of net transport rather than individual channel transport.
For $d/a_1 = 1.47$, where NDC of $V_1$ appeared in Fig.~\ref{fig:vel2}(b),
the inset of Fig.~\ref{fig:vel2}(d) shows that there is no NDC in the 
overall velocity $V$.
This is because the decrease in $V_1$ at $F^D=0.36$ is exactly compensated
by the increase in $V_2$, so there is no net drop in $V$ at the second
depinning transition.
For both $d/a_1=1.47$ and $d/a_1=1.64$, $V$ shows a characteristic step
feature that indicates the presence of a two-step depinning transition.
On the lower step in $V$, only channel 1 is depinned, while on the upper step,
both channels are flowing.
For this particular set of parameters,
we find only a weak signature in $V$ of the relocking transition that occurs
at higher drives;
however, for other parameters, we will show below that the onset of
dynamical locking produces much more pronounced effects in $V$.

\begin{figure}
\includegraphics[width=3.5in]{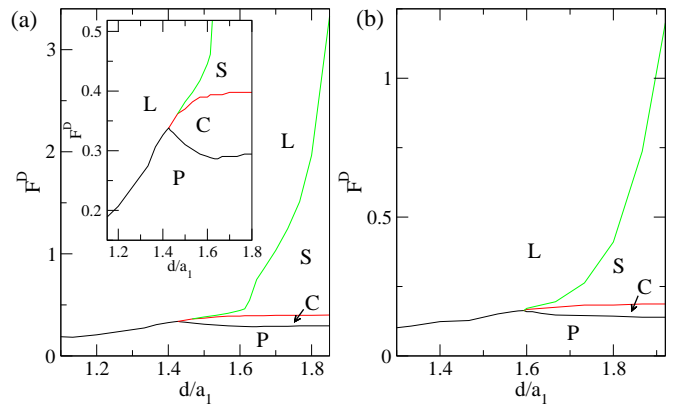}
\caption{
Dynamic phase diagrams $F^D$ vs $d/a_1$ for the $M=2$ system from 
Fig.~\ref{fig:vel2}
with $N_1=N_2$, $n_1=n_2$, $n_1/N_1=0.25$, $R_p=0.23a_0$, and $a_1=a_2=1.5a_0$.
P: pinned phase with $V_1=V_2=0$; C: coexistence phase where one channel
is moving but the other is pinned; S: sliding phase where both channels are
moving but $V_1 \neq V_2$; L: dynamically locked phase where $V_1=V_2>0$.
(a) The sample from Fig.~\ref{fig:vel2} with $F^{p}_1 = F^p_2 = 6.0$. 
For $d/a_1 < 1.43$ the system depins directly into the locked phase, while
for $d/a_1 > 1.43$, dynamically induced locking occurs at a value of $F^D$
that increases with increasing $d/a_1$.
Inset: A blowup of the main panel illustrating that the P phase reaches its
maximum value of $F^D$ at the transition from elastic to plastic depinning.
(b) A sample with $F^{p}_1 = F^p_2 = 3.0$
has the same features but the onsets of the S and C phases fall at
higher values of $d/a_1 >1.6$. 
}
\label{fig:phase3}
\end{figure}

By conducting a series of simulations we map the dynamic phase diagram
for $F^{D}$ vs $d/a_1$ for the sample from Fig.~\ref{fig:vel2} with
$F^p_1=F^p_2=6.0$ in Fig.~\ref{fig:phase3}(a) and for a sample with 
$F^p_1=F^p_2=3.0$ in
Fig.~\ref{fig:phase3}(b).
In the pinned (P) phase, both channels are pinned, $V_1=V_2=0$.
In the coexistence (C) phase, 
one channel is pinned while the other is moving. 
The sliding (S) phase has both channels moving but not locked,
$V_1>0$ and $V_2>0$ but $V_1 \neq V_2$,
while in the locked (L) phase both channels move together without
slipping, $V_1=V_2>0$.
In Fig.~\ref{fig:phase3}(a), for $d/a_1 > 1.43$ the layers decouple
at depinning and the width of region S grows  
with increasing $d/a_1$. 
The width of the coexistence phase saturates as $d/a_1$ increases due to
the decoupling of the particles, which causes the depinning thresholds of
each channel that mark the borders of the C phase 
to be determined only by the quenched disorder configuration and the 
interactions among particles within the channel, and to be insensitive to
the positions of the particles in the neighboring channel.
The recoupling that marks the transition between the S and L phases
rapidly shifts to higher $F^D$ with increasing $d/a_1$ for
$d/a_1 >  1.613$. 
In Fig.~\ref{fig:phase3}(b) we show that in a sample with weaker 
pinning of $F^p_1=F^p_2=3.0$, the transition directly from P to L persists
out to the higher value of $d/a_1=1.6$ but that the same general
features found for higher $F^p$ in Fig.~\ref{fig:phase3}(a) still occur
and are shifted to higher $d/a_1$ and lower $F^D$.

The dynamical phase diagrams in Fig.~\ref{fig:phase3} 
have many similarities to the dynamical phase diagrams obtained for 
2D and 3D driven vortex systems \cite{Moon,MC,Saunders,Olson,OlsonN}. 
The 2D vortex system exhibits pinned, plastic, and dynamically
ordered phases, with the vortex lattice ordering into either a moving
smectic or moving crystal configuration 
\cite{Moon,Olson,Koshelev}. 
Our system has the same pinned phase, while the coexistence phase that
contains a mixture of moving and pinned channels corresponds to the plastic
flow regime for the vortex system.
The sliding phase we observe resembles 
the moving smectic phase found for vortices, 
while our locked phase is equivalent to the 2D dynamically reordered phase
or, for 3D vortex systems, to the transition from decoupled layers to
coupled 3D vortex lines
\cite{Zimanyi,Scalettar,Zhao}.
In the vortex system,
as the pinning strength increases or the vortex-vortex interactions are
weakened, the reordering transition shifts to higher drives \cite{Moon}.
The dynamical reordering occurs because the pinning is effectively weakened
when the vortices are moving rapidly.
In our system, 
in the locked phase the effectiveness of the quenched disorder
is dramatically reduced.
As the interchannel particle-particle interaction strength decreases
for increasing $d/a_1$, the quenched disorder becomes more effective,
the recoupling transition shifts to higher $F^D$, and the additional
C and S phases appear between the P and L phases.
  
In the inset of Fig.~\ref{fig:phase3}(a) 
we plot a blowup of the region near $d/a_1=1.4$ for the $F^p_1=F^p_2=6.0$ sample
where the transition between elastic depinning directly into the locked phase
and plastic depinning into the coexistence phase occurs.
The critical
depinning threshold $F_c$ reaches a peak at 
$d/a_1=1.43$ at the transition, while
for $d/a_1>1.43$, $F_c$ decreases and then saturates with increasing $d/a_1$.
The $F_c$ peak 
has all the hallmark features 
of the peak effect phenomenon found for vortices and 
other systems where a peak in the critical depinning threshold occurs
near the transition from elastic to plastic depinning
\cite{Higgins,Kes,Olson,Menon}. In many of these studies the $F_c$
peak occurs inside the plastic depinning regime
just beyond the transition point, while $F_c$ for the elastic depinning
is always lower than for the plastic depinning \cite{Higgins}. 
In our two layer system, the $F_c$ peak falls
right at the transition out of the elastic depinning regime,
and $F_c$ decreases with increasing $d/a_1$ within the 
plastic depinning regime. 

\begin{figure}
\includegraphics[width=3.5in]{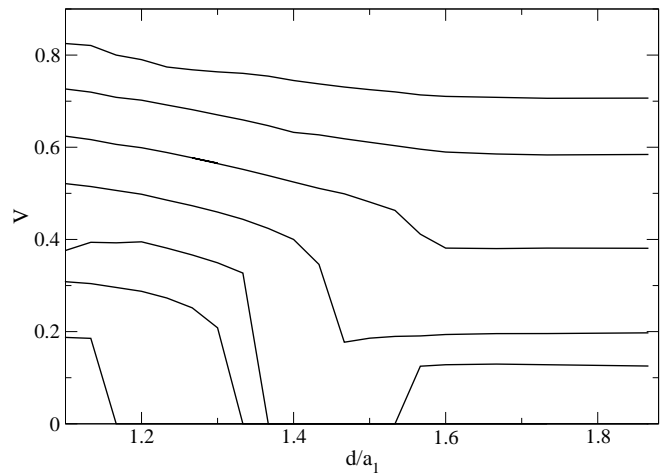}
\caption{
The summed velocity $V=V_1+V_2$ for the $M=2$ 
system in Fig.~\ref{fig:phase3}(a) with $N_1=N_2$, $n_1=n_2$, $n_1/N_1=0.25$,
$R_p=0.23a_0$, $a_1=1.5a_0$,
and $F^p_1=F^p_2=6.0$ 
at $F^{D} = 0.4$, 0.5, 0.6, 0.7, 0.8, 0.9, and $1.0$,
from bottom to top. 
The value of $V$ in the decoupled regime is always 
lower than in the coupled regime, and the dip in $V$ at 
the decoupling transition
gradually fades away as $F^D$ increases.
}
\label{fig:avgv5}
\end{figure}

Determining whether the effectiveness of the pinning in the plastic or
decoupled phases is higher than in the elastic phase
can be ambiguous when different quantities are measured \cite{Higgins,Kes}.
If we consider only the overall velocity $V$ at a 
fixed value of $F^D$ as we vary $d/a_1$,
we find that $V$ drops in the decoupled regime, suggesting that the
pinning is more effective in the decoupled state.
This contrasts with the behavior of the depinning threshold $F_c$, which
drops in the decoupled regime, suggesting that the pinning is less
effective in the decoupled state. 
In Fig.~\ref{fig:avgv5}, we plot $V$
versus $d/a_1$ for the sample from Fig.~\ref{fig:phase3}(a) with 
$F^p_1=F^p_2=6.0$ for
fixed $F^D$ values ranging from $F^D=0.4$ to $F^D=1.0$.
For $F^D\leq 0.8$, V drops or becomes zero at the decoupling transition,
and for $0.6 \leq F^D \leq 0.8$ the value of $V$ 
for $d/a_1$ above the decoupling 
transition is smaller than the value of $V$ 
for $d/a_1$ below the decoupling transition.
For $F^D>0.8$, the 
size of the dip in $V$ at the decoupling transition gradually
diminishes.
This behavior is 
very similar to that observed across the peak effect for vortices
in type-II superconductors, 
and indicates that for the moving particles, the pinning is more effective
in the decoupled phase than in the coupled phase.
The behavior of $V$ differs from that of $F_c$ because $V$ is associated with
moving particles while $F_c$ is determined by the static configuration of
the pinned particles.

\begin{figure}
\includegraphics[width=3.5in]{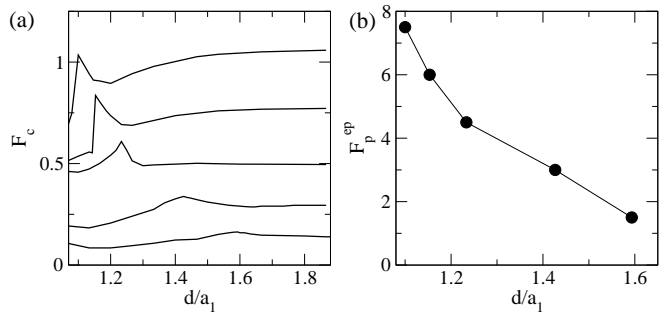}
\caption{
(a) The critical depinning threshold $F_{c}$ vs $d/a_1$ for the $M=2$ system
in Fig.~\ref{fig:phase3}(a) 
with $N_1=N_2$, $n_1=n_2$, $n_1/N_1=0.25$, $R_p=0.23a_0$, and $a_1=1.5a_0$
for varied $F^{p}_1=F^p_2 = 15.0$, 12.0, 9.0, 6.0, and $3.0$, from top to bottom. 
In each case
the peak in $F_{c}$ falls at the transition from elastic to 
plastic depinning. 
(b) $F_p^{ep}$, the value of $F^{p}_1$ and $F^p_2$ at 
which a transition from elastic to plastic
depinning occurs, vs $d/a_1$ in the same system.
}
\label{fig:fc5}
\end{figure}

In Fig.~\ref{fig:fc5}(a) 
we plot $F_c$ versus $d/a_1$ for a range of $F_p$ values in the
system from Fig.~\ref{fig:phase3}(a)
to illustrate
that the decoupling transition is associated with a
peak in $F_{c}$.
On average, $F_{c}$ increases with increasing $F^p$ and
with increasing $d/a_1$.  The decoupling transition, and 
with it the peak in $F_c$, shifts to 
lower values of $d/a_1$ with increasing $F^{p}$ and becomes more prominent. 
In all cases the peak in $F_{c}$ falls exactly at
the transition from elastic to plastic flow.
For a given value of $d/a_1$, it is possible to induce a decoupling
transition by increasing $F^p$; the value of $F^p_1$ and $F^p_2$ at the 
decoupling transition is termed $F_p^{ep}$.
In Fig.~\ref{fig:fc5}(b) we plot $F_{p}^{ep}$ 
versus $d/a_1$, where we find that as 
$d/a_1$ decreases and the coupling between the channels
increases, $F_p^{ep}$ increases faster than linearly.

\begin{figure}
\includegraphics[width=3.5in]{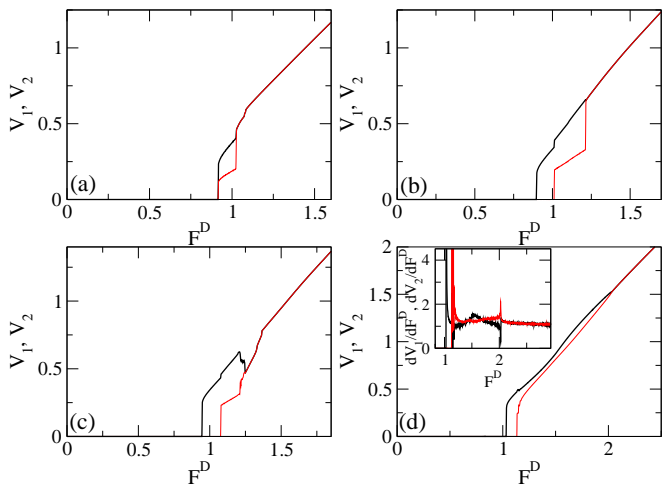}
\caption{
$V_{1}$ (dark line) and $V_{2}$ (light red line) vs $F^{D}$ for the 
$M=2$ system in Fig.~\ref{fig:fc5}(a) with $N_1=N_2$, $n_1=n_2$, 
$n_1/N_1=0.25$, $R_p=0.23a_0$, $a_1=1.5a_0$,
and $F^p_1=F^p_2=15.0$ for
(a) $d/a_1 = 1.147$,
(b) $d/a_1 = 1.2$, 
(c) $d/a_1 = 1.27$, and 
(d) $d/a_1 = 1.47$. 
Inset of (d): The $dV_1/dF^D$ and $dV_2/dF^D$ vs $F^D$ curves both peak
at the dynamical coupling transition at $F^D=2.0$; for $F^D>2.0$ above
the transition, the fluctuations in both curves show increased
correlation.
}
\label{fig:v6}
\end{figure}

The velocity force signatures illustrated in Fig.~\ref{fig:vel2} are
generally robust against changes in pinning strength.
For samples with stronger pinning, $F^p_1=F^p_2 >6.0$,  many of the features become
more prominent.  
For sufficiently strong $F^p$, the system can 
depin directly into the sliding state with no intermediate coexistence
state, as illustrated in
\ref{fig:v6}(a) for $F^{p}_1=F^p_2 = 15.0$ and $d/a_1 = 1.147$. 
Here both channels depin simultaneously at $F^{D} = 0.915$, and 
for $0.915 < F^{D} < 1.025$ the channels move at different velocities with
$V_1 \neq V_2$. 
We also find a larger number of jumps in the 
velocity force curves for higher $F^{p}$, 
as shown in Fig.~\ref{fig:v6}(a) near $F^{D} = 1.08$. 
The jumps are indicative of the solitonlike nature of the particle motion
in channels with strong pinning below the recoupling transition.  
Due to the random spatial placement of the
pinning sites, the individual pins vary in their effectiveness at trapping
particles.  As the driving force is increased below depinning, particles
escape from the least effective pins only to pile up behind particles that are
trapped in the more effective pins.  Since the channel is 1D, flowing
particles cannot pass trapped particles.  As the driving force further
increases, particles in the pileup regions approach each other more closely
and exert a greater force on the trapped particle that produced the
pileup.  At depinning, the trapped particle escapes from the pin and is
immediately replaced by another particle that becomes pinned in its place.
The trapped particle travels across the system and rejoins the pileup; the
extra force it exerts on the particles ahead of it causes the trapped particle
at the front of the pileup to depin and repeat the process.  This picture
is oversimplified; in actuality, each channel contains multiple 
trapping sites that create pileups and particles
jump from one pileup to another above depinning.  Each of the trapping sites
is associated with some local depinning threshold $F_c^{loc}$.  As $F^D$
increases, one trapping site after another reaches the condition
$F_c^{loc}<F^D$ and ceases to trap particles.  The resulting enhancement of
the mobility of all the particles in the channel manifests itself as a jump
in channel velocity $V_j$.  As $F^p$ increases, there is a greater spread
in $F_c^{loc}$, producing a larger number of steps in $V_1$ and $V_2$ above
depinning.
We note that such velocity force jumps
are often called switching events in 
sliding charge density wave systems 
\cite{Fisher}.
For vortex systems, near the peak
effect regime a series of jumps and dips in the current-voltage curves
and $dV/dI$ curves can appear.  These have
been termed a fingerprint phenomenon since the same features 
repeat upon cycling the IV curves, indicating that the features result from
the detailed configuration of the pinning sites \cite{SB,Dominguez}. 
The curves in Fig.~\ref{fig:v6}(a) are obtained for
a specific pinning realization. 
For a different pinning realization, the general features of the velocity
force curves remain the same but the location and height 
of the velocity jumps will change.
In Fig.~\ref{fig:v6}(b), a sample with 
the same disorder configuration and strength but with $d/a_1 = 1.2$ 
has separate depinning thresholds for channels 1 and 2
along with a sharp transition into the locked phase at $F^D=1.2$.
In the $d/a_1= 1.27$ sample of Fig.~\ref{fig:v6}(c), 
we observe a strong negative differential conductivity signature in $V_1$
at the transition between the sliding and locked phases.
There is also a change in the rate at which $V_1$ and $V_2$ increase
with $F^D$ in the locked phase, with a more rapid increase
for $1.25 < F^{D} < 1.365$ 
and a slower increase for $F^{D} > 1.365$. 
For $d/a = 1.47$ in Fig.~\ref{fig:v6}(d), 
the channels do not lock until
$F^{D} = 2.03$, and $V_{1}$ shows additional structure within the sliding
phase.
For increasing $F^{p}$ beyond what we show here, we find
similar features in the velocity force curves; however, 
the number of jumps and anomalies in the curves further increases
as the microscopic configurations of  
the pinning sites begin to dominate
the behavior completely.

\begin{figure}
\includegraphics[width=3.5in]{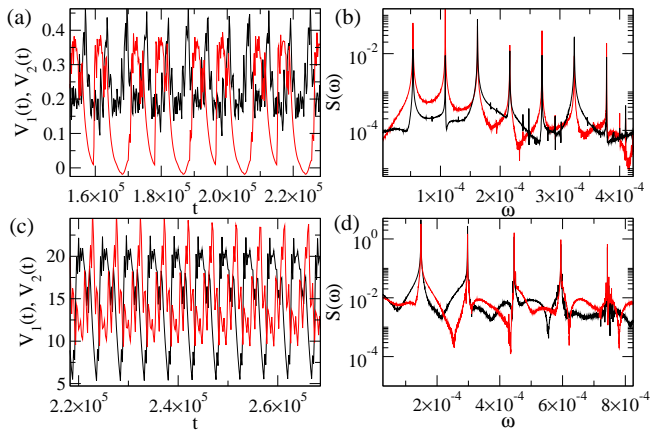}
\caption{
(a) The instantaneous velocity of each channel $V_1(t)$ and $V_2(t)$ versus
time in simulation steps for channels 1 (dark black line) and 2 
(light red line) for the $M=2$ system in 
Fig.~\ref{fig:vel2}(c) 
with $N_1=N_2$, $n_1=n_2$, $n_1/N_1=0.25$, $R_p=0.23a_0$, $a_1=1.5a_0$,
$d/a_1=1.57$, and $F^p_1=F^p_2=6.0$
at $F^{D} = 0.395$.  The system is in the S phase
and both channels are moving at different velocities. 
(b) $S(\omega)$ for the time series
in panel (a) shows that channel 2 has greater spectral power at the lowest
frequencies compared to channel 1.  $\omega$ is reported in inverse
simulation time steps.
(c) $V_1(t)$ and $V_2(t)$ for the same system with
$F^{D} = 0.47$ where the channels are locked. 
(d) $S(\omega)$ for the time series in panel (c) shows nearly
equal spectral weight in each channel at the fundamental frequencies.
}
\label{fig:sofw8}
\end{figure}

We next compare the velocity fluctuations in the individual
channels in the unlocked and locked state.
Fig.~\ref{fig:sofw8} shows the instantaneous velocities
$V_1(t)$ and $V_2(t)$ for channels 1 and 2 as a function of time measured
in simulation steps
for the system in Fig.~\ref{fig:vel2}(c) at $d/a_1 = 1.57$ with 
$F^{p}_1 = F^p_2= 6.0$.
At $F^{D} = 1.395$ in Fig.~\ref{fig:sofw8}(a), the system is in the S phase
and both channels are moving with different average velocities $V_1>V_2>0$.
Both channels show a periodic response but channel 1 has a higher 
frequency and higher average velocity, while
$V_2(t)$ drops nearly to zero during each cycle.
In Fig.~\ref{fig:sofw8}(b) we plot the corresponding power spectra $S(\omega)$ 
of the velocity signals, with $\omega$ measured in inverse
simulation time steps.  Channel 2 has greater spectral weight at the low
frequency of $\omega=1.1 \times 10^{-4}$ corresponding to its 
fundamental frequency,
while the peak at this frequency for channel 1 is orders of
magnitude smaller in height.  Channel 1 shows a response at this frequency
due to its coupling with channel 2.
Near $\omega=1.62 \times 10^{-4}$, the fundamental frequency of
the faster moving channel 1, the peak in $S(\omega)$ is higher for
channel 1 than for channel 2.
This result shows that the two channels each have a periodic velocity
signal corresponding to a different washboard frequency; however, since
the channels are interacting, both velocity signals contain both
washboard frequencies.
In Fig.~\ref{fig:sofw8}(c) we plot $V_1(t)$ and $V_2(t)$ for the same system
at $F^{D} = 0.47$ in the locked regime where the channels
have the same average velocity. 
Here both channels exhibit the same fundamental frequency even though
the shapes of $V_1(t)$ and $V_2(t)$ do not completely overlap.
The corresponding $S(\omega)$ in Fig.~\ref{fig:sofw8}(d) shows that both 
channels have nearly the same spectral weight at the fundamental frequency
of $\omega=1.48 \times 10^{-4}$ and its harmonics, unlike the unequal peaks which
appeared in the unlocked phase in Fig.~\ref{fig:sofw8}(b).

\begin{figure}
\includegraphics[width=3.5in]{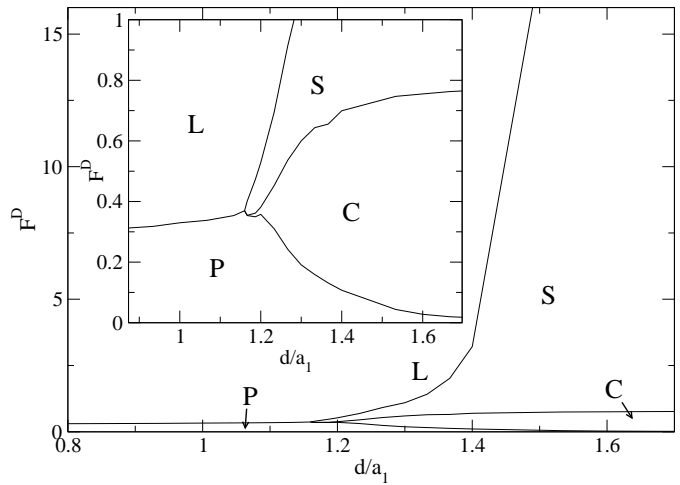}
\caption{
The dynamic phase diagram $F^D$ vs $d/a_1$ for the $M=2$ system in 
Fig.~\ref{fig:vel2} with
$N_1=N_2$, $n_1=n_2$, $n_1/N_1=0.25$, $R_p=0.23a_0$,
$a_1=1.5a_0$,
$F^{p}_1 = 0$ and $F^{p}_2 = 15.0$.
P: pinned phase; C: coexistence phase where only one channel is moving;
S: sliding phase where $V_1 \neq V_2 > 0$; L: locked phase.
As $d/a_1$ increases, the depinning threshold for channel 1
goes to zero and the system 
immediately enters region C for small nonzero $F^D$.  
Inset: A blowup of the main panel near
the transition from elastic to plastic depinning. 
}
\label{fig:phase9}
\end{figure}

\section{Varied Relative Disorder Strength and Hysteresis}

We next consider the effects of varying the pinning strength from channel
to channel in order to create one channel with high pinning strength and 
one channel with low pinning strength.
Figure~\ref{fig:phase9}(a) 
shows the dynamic phase diagram of $F^{D}$ versus $d/a_1$ 
for a two channel system with 
$F^{p}_1 = 15.0$ 
and $F^{p}_2 = 0$, while the inset illustrates a blowup of the
area near the onset of plastic depinning.
The strongly pinned particles in channel 1 are able to effectively pin the
particles in channel 2 via particle-particle interactions alone, allowing
the pinned regime to persist for finite values of $F^D$  
in spite of the fact that the pins in channel 2 have zero strength.
For low $d/a_1$ when the interactions between particles in neighboring
channels are strong,
the depinning of both channels occurs simultaneously
in a single step at $F_c$ from
the pinned state to the moving locked state. 
As $d/a_1$ increases, 
$F_{c}$ increases and reaches a maximum value at $d/a_1 = 1.16$ corresponding
to the decoupling transition.
For $d/a_1$ above the decoupling transition, 
the depinning force $F_{c}$ for channel 2
decreases monotonically with increasing $d/a_1$, unlike the saturation
that occurred in Fig.~\ref{fig:phase3} for samples with $F^p_1=F^p_2$.
As $d/a_1$ increases, the coupling between the channels decreases
monotonically, and since the particles in channel 2 are pinned only due to
their interactions with the pinned particles in channel 1, 
$F_c$ for channel 2 decreases with increasing $d/a_1$.
The transition line between the C and S phases, 
which corresponds to the depinning force $F_c$ for channel 1,
grows more rapidly 
above the decoupling transition
with increasing $d/a_1$ in Fig.~\ref{fig:phase9} than it
did for samples with $F^p_1=F^p_2$ in Fig.~\ref{fig:phase3}.
This is because the particles in channel 2 produce relatively
little effective drag for the particles in channel 1 when there is no
pinning in channel 2, and this drag rapidly becomes nearly zero as
$d/a_1$ increases and the coupling between the channels decreases.
The transition from the S to L phases also grows more rapidly with 
increasing $d/a_1$ in Fig.~\ref{fig:phase9} than for 
the samples with $F^p_1=F^p_2$.
Our results show that coupling-decoupling transitions can persist
even when the pinning in one channel is completely absent.  

\begin{figure}
\includegraphics[width=3.5in]{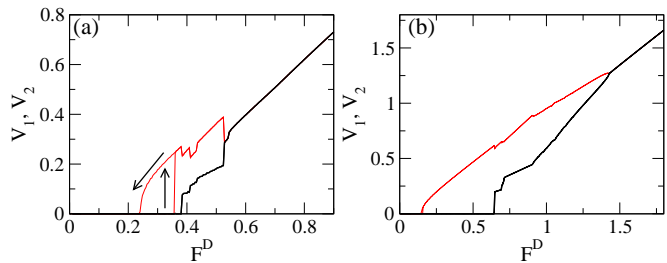}
\caption{
$V_1$ (dark line) and $V_2$ (light red line) vs $F^D$ under a cycled
drive for the $M=2$ system in Fig.~\ref{fig:phase9} with 
$N_1=N_2$, $n_1=n_2$, $n_1/N_1=0.25$, $R_p=0.23a_0$, $a_1=1.5a_0$,
$F^p_1=15.0$ and $F^p_2=0.0$.
(a) At $d/a_1 = 1.27$ we find hysteresis in $V_1$ near the 
transition to the pinned phase but there is no hysteresis in $V_2$.
The arrows indicate the $V_2$ curves obtained for sweeping $F^D$ up and down.
(b) For the weaker coupling of $d/a_1=1.33$, there is no hysteresis and
dynamic coupling occurs at $F^{D} = 1.43.$ 
}
\label{fig:hyst10}
\end{figure}

For the parameters examined so far in this work we observe only 
weak hysteresis in the velocity-force curves 
occurring near the sharp jumps in $V_1$ or $V_2$.
Hysteresis is often absent in 1D systems, as shown with
the no-passing rule \cite{Mid}; however,
in studies using phase-field modeling,
hysteresis has been observed in 1D systems \cite{Elder}. 
In our system
we never find hysteresis when the depinning is elastic, but 
hysteresis can occur for plastic depinning, as illustrated
in Fig.~\ref{fig:hyst10}(a), for a system with the
same parameters as in Fig.~\ref{fig:phase9} at 
$d/a_1 = 1.27$. 
Here hysteresis occurs at the depinning transition only for the pin-free
channel 2 but not for channel 1.
The summed velocity $V=V_1+V_2$ also exhibits hysteresis at depinning,
but by examining the separate velocity signals we can determine
that the hysteresis originates from only one of the channels.
Figure~\ref{fig:hyst10}(a) also shows that there is no 
hysteresis across the S-L transition at $F^D=0.526$ 
in spite of the discontinuous 
jump in both $V_1$ and $V_2$ that occurs at this transition.
In Fig.~\ref{fig:hyst10}(b) we plot $V_1$ and $V_2$ 
for cycled $F^D$ in a more weakly coupled sample with $d/a_1=1.33$,
further from the value of $d/a_1=1.16$ where the transition from elastic to
plastic depinning occurs. 
In this case we find dynamic locking 
of the channels above $F^{D} = 1.43$; 
however, there is no hysteresis across any of the transitions. 

\begin{figure}
\includegraphics[width=3.5in]{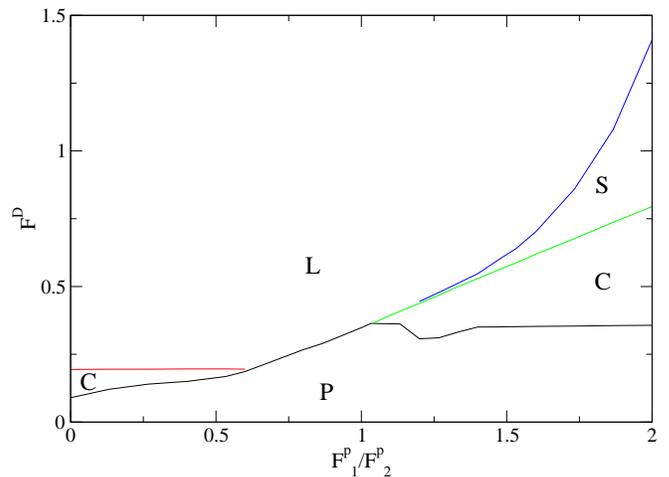}
\caption{
Dynamic phase diagram $F^D$ vs $F^p_1/F^p_2$ for an 
$M=2$ system with $N_1=N_2$, $n_1=n_2$, $n_{1}/N_1 = 0.182$,
$R_p=0.35a_0$, $a_1=1.5a_0$,
$d/a_1 = 1.7$, and fixed $F^p_2=0.75$.
P: pinned phase; C: coexistence phase; S: sliding phase; L: locked phase.
Near $F^p_1/F^p_2 = 1.0$, where the pinning strength 
is the same in both channels, the system depins elastically. 
The depinning becomes plastic once $F^p_1/F^p_2$ moves sufficiently far
away from 1 in either direction.
}
\label{fig:phase11}
\end{figure}

We can also examine the effects of varying the 
relative pinning strength in the two channels
by altering $F^p_1/F^p_2$
for a system with $N_1=N_2$, $n_1=n_2$, 
$n_1/N_1 = 0.182$, $R_{p} = 0.35a_0$, 
and 
$d/a_1 = 1.7$ as shown in Fig.~\ref{fig:phase11}. 
Here $F^{p}_{2} = 0.75$
and $F^{p}_{1}$ is varied. 
For this particular set of parameters at
$F^{p}_{1}/F^{p}_{2} = 1.0$ the depinning is elastic.  
For 
$0 < F^{p}_{1}/F^{p}_{2} < 0.6$, 
the system depins plastically into the coexistence state where channel 1 is
moving and channel 2 is pinned, and for higher $F^D$ the system enters
the locking regime without passing through the sliding phase.
At $F^{p}_{1}/F^{p}_{2} = 0$ the depinning threshold of channel 1 is
nonzero due to the strong interaction of the particles in channel 1 with the
pinned particles in channel 2.
As $F^p_1$ increases from zero the width of the pinned region grows
as the depinning threshold of channel 1 rises;
however, for $0<F^p_1/F^p_2<0.6$ the depinning threshold of channel 2
remains constant since $F^p_2$ is fixed.
For $ 0.6 < F^{p}_{1}/F^{p}_{2} < 1.0$ the depinning is elastic
and both channels depin simultaneously; further, within this regime
the pinned region has the most rapid growth for increasing
$F^p_1/F^p_2$. 
For $F^{p}_{1}/F^{p}_{2} > 1.0$ the depinning becomes plastic again and each
channel depins separately; however, channel 2 now depins first
since $F^{p}_{1} > F^{p}_{2}$.  
As $F^{p}_{1}/F^{p}_{2}$ increases above 1, 
the width of the pinned phase passes through a small dip
but remains nearly constant since the depinning threshold of channel 2
is determined by $F^{p}_{2}$ which is held fixed. 
Near $F^p_1/F^p_2=1.2$ we find the onset of the sliding phase, which grows
rapidly in width with increasing $F^p_1/F^p_2$. The sliding phase is followed at
higher $F^D$ by the dynamically locked phase.
If we increase the fixed value of $F^p_2$ and sweep $F^p_1/F^p_2$,
the extent of the elastic depinning window diminishes as $F^p_2$ becomes
larger until for sufficiently large $F^p_2$ there is no longer an elastic
depinning regime.  Instead, the sliding phase appears for all values
of $F^p_1/F^p_2$.
We have also considered samples with $F^{p}_1=F^p_2$  but with 
different pinning densities $n_1 \neq n_2$, and we find the same general
features as a function of $n_1/n_2$ as we have shown here for varying
$F^p_1/F^p_2$.

\begin{figure}
\includegraphics[width=3.5in]{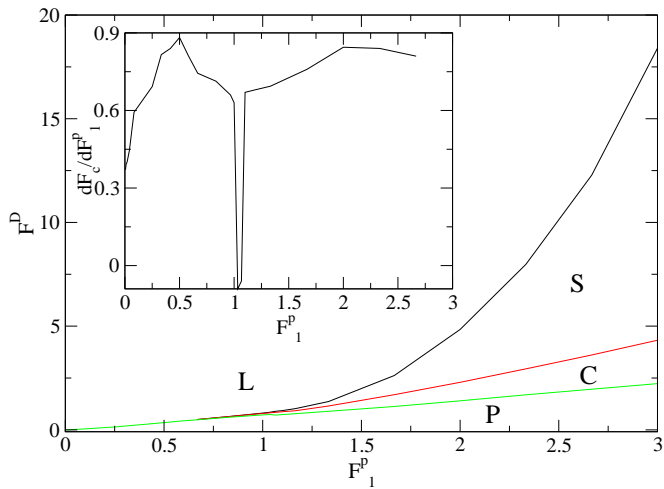}
\caption{
Dynamic phase diagram $F^{D}$ vs $F^{p}_1$ for an $M=2$ sample with 
$F^p_2=F^p_1$, $N_1=N_2$, $n_1=n_2$, 
$n_1/N_1=0.25$, 
$R_p=0.23a_0$,
$a_1=1.5a_0$,
and $d/a_1=1.7$.
P: pinned phase; C: coexistence phase; S: sliding phase; L: locked phase.
The depinning
is elastic for $F_{p} < 0.8$.
Inset: $dF_c/dF^{p}_1$ 
for the same system shows a pronounced dip
near the transition from elastic to plastic depinning.
}
\label{fig:phase12}
\end{figure}
           
In Fig.~\ref{fig:phase12} we 
plot the dynamic phase diagram for $F^{D}$ versus 
$F^{p}_1$ in a sample with equal pinning strength in 
both channels, $F^p_2=F^p_1$, and with
$d/a_1 = 1.7$. 
For $ F^{p}_1 <  0.7$, the depinning
is elastic and the
depinning force $F_{c}$ drops to zero as $F^{p}_1$ drops to zero. 
For $F^{p}_1 > 0.8$, the depinning is plastic 
and all of the dynamical transition lines shift to higher $F^D$
for increasing $F^{p}_1$, with the S to L transition 
increasing the most rapidly. 
In the inset of Fig.~\ref{fig:phase12} we plot
$dF_{c}/dF^{p}_1$ versus $F^p_1$ where we find a roughly linear increase
in the regime $0 < F^{p}_1 < 0.55$. 
This implies that $F_c \propto (F^{p}_1)^2$, which is consistent with
elastic depinning in the collective regime \cite{Moon}.
Just above the transition to plastic depinning, 
$dF_{c}/dF^{p}_1$ passes through a strong drop and then
recovers to a nearly constant value near 1,
consistent with $F_{c} \propto F^{p}$ as expected for the single particle limit. 
This results 
indicates that the transition from elastic to plastic
depinning is associated with clearly observable features in the
transport curves.
If we fix $F^{p}_1$ and instead increase the density of pinning sites 
while holding $n_1=n_2$, we find dynamical phases with tendencies similar to
those shown in Fig.~\ref{fig:phase12}.

\begin{figure}
\includegraphics[width=3.5in]{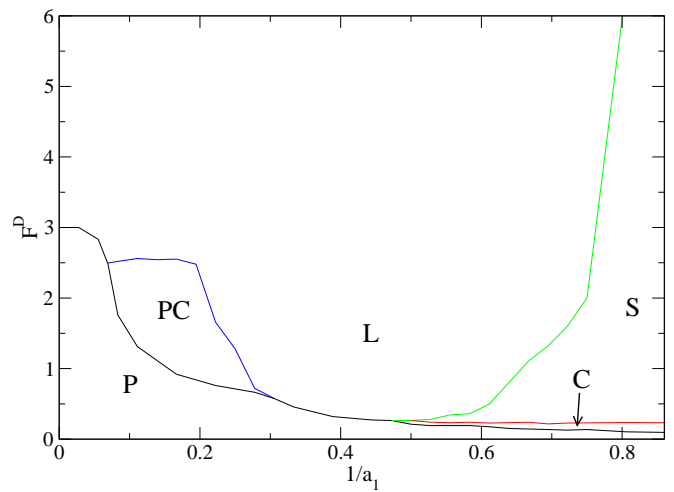}
\caption{
Dynamic phase diagram $F^{D}$ vs density $1/a_1$
for an $M=2$ sample with
$N_1=N_2$, $n_1=n_2$, $a_1=a_2$, $n_1/N_1=0.25$,
$R_p=0.35a_0$, $d=2.75a_0$, and $F^p_1=F^p_2=3.0$. 
P: pinned phase; C: coexistence phase; S: sliding phase; L: locked phase.
For $0.07 < 1/a_1 < 0.305$, the system transitions from a pinned phase
into a pulsed coexistence phase PC.
}
\label{fig:phase13}
\end{figure}

\section{Density Dependence and Dynamic Commensuration Effects}

We next consider the effect of holding the distance $d$ between channels fixed
while varying the density of the particles which changes the
lattice constants $a_1$ and $a_2$. In Fig.~\ref{fig:phase13} 
we plot the dynamic phase
diagram as a function of $F^D$ and density $1/a_1$ for varied $a_1=L/N_1$ in 
a system with an equal number of particles
in each channel, $N_1=N_2$, $F^p_1=F^p_2=3.0$,
and with fixed $d=2.75$.
The depinning threshold $F_c$ generally decreases with increasing 
density
due to the increase in the particle-particle interactions relative
to the pinning strength as the particles get closer together within 
each channel.  
For intermediate densities of $0.305 < 1/a_1 < 0.47$, 
the depinning is elastic, while     
for $0.07 < 1/a_1 < 0.305$, 
the depinning is plastic and the sample enters either the coexistence phase C,
the sliding phase S, or a pulsed coexistence phase PC.
Each channel is alternately pinned or moving in the PC phase, so that this
phase resembles the C phase except that the stationary and moving channels
keep switching places.
At lower densities $1/a_1<0.305$, the system no longer depins elastically
due to the increase of the effectiveness of the pinning, which induces a
decoupling of the channels.
For $1/a_1<0.07$, we have $N_1>n_1$.  Since there are more pinning sites than
particles, the depinning occurs in the single particle limit and it is
no longer meaningful to describe the pinning as plastic or elastic since the
particles in the channels are essentially noninteracting at depinning.
Additionally, since $F^p_1$ and $F^p_2$ are held fixed, at these low
densities all the particles in each channel depin simultaneously and move
at the same velocity so the C and S phases cannot occur.

Figure~\ref{fig:phase13} shows that
for $1/a_1 > 0.47$, the depinning is plastic again and
the S-L transition shifts to higher $F^D$ 
with increasing $1/a_1$. This might 
seem surprising since at higher densities the interactions between
particles in the same channel become stronger;
however, it is known from studies
of layered vortex systems that increasing the field can reduce the
coupling {\it between} layers for fields well 
below $H_{c2}$ \cite{Scalettar}. 
As the density increases, the interactions between particles in the same
channel become stronger more rapidly than the interactions between particles
in neighboring channels.  This enhances the S phase since it 
becomes more difficult
for the channels to lock.
In order to minimize energy, the moving particles 
adopt a zig zag configuration with the particles in one channel 
offset relative to the particles in the other channel.  This approximates a
triangular lattice configuration.  As $1/a_1$ increases, the lattice
spacing along the channels shrinks but the lattice spacing perpendicular
to the channels is fixed, producing an effectively increasingly anisotropic
triangular lattice configuration.  Eventually the spacing becomes so 
anisotropic that there is very little energy difference between an
alternating configuration and one in which the particles are adjacent to
each other, so the channels decouple.

The phase diagram in Fig.~\ref{fig:phase13} 
exhibits a number of features found in phase diagrams 
for strongly layered superconductors
containing disordered pinning sites. 
For example, the superconducting system undergoes
a low field decoupling transition when the pancake vortices in
each layer are far apart and only weakly interacting, 
while at higher fields there is a field induced  
decoupling \cite{Scalettar}. 
In experiments, samples show both a low field disordering transition 
and a high field decoupling transition \cite{Banjerjee}.
It was argued that the low field transition is pinning induced and occurs
when the pinning energy overwhelms the weak vortex-vortex interactions at
low vortex density, 
while the higher field disordering transition indicates the onset of
the peak effect \cite{Banjerjee}.
In our system, as we increase $F^{p}$ the lower and upper endpoints of the
elastic depinning regime approach each other until the elastic depinning
disappears completely.
Conversely as we decrease $F^p$ 
the elastic depinning region becomes more extended.  
Although there is a small increase 
in $F_c$ at the lower transition from plastic to
elastic depinning, we find no feature in $F_c$ at the upper transition from 
elastic to plastic depinning; however, if we measure the average particle
velocity for fixed drive,
the velocity drops across the higher density transition, 
similar to the effect illustrated in Fig.~\ref{fig:avgv5}.

\begin{figure}
\includegraphics[width=3.5in]{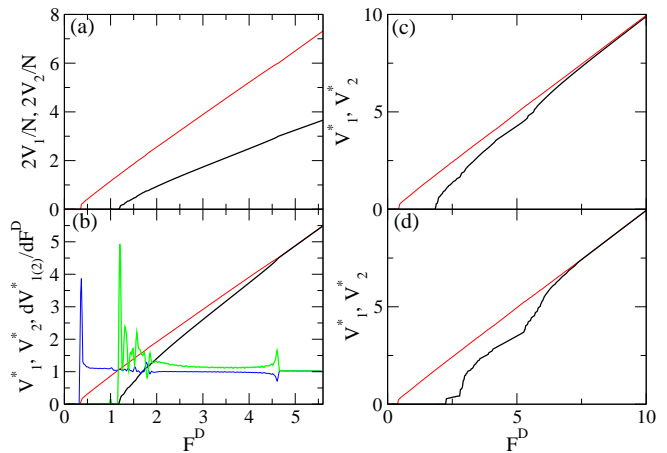}
\caption{
Channel velocities vs $F^D$ for an $M=2$ system
with $F^p_1=F^p_2=6.0$, $n_1=n_2$, $n_2/N_2=0.167$, 
$R_p=0.23a_0$, and $d/a_2=1.33$ for fixed $N_2$ and varied $N_1$.
(a) $2V_1/N$ (lower dark line) and $2V_2/N$ (upper light red line) for
$N_{1}/N_{2} = 1/2$.  The velocities are normalized by $N/2$, where
the total particle number $N=N_1+N_2$.
Channel 2 contains a larger number of particles than channel 1 and therefore
$V_2$ increases more rapidly than $V_1$.
(b) The same curves plotted as $V^*_1=V_1/N_1$ (lower dark line) and 
$V^*_2=V_2/N_2$ (upper light red line).
Here a clear dynamical locking transition appears.
Also shown are the corresponding
$dV^*_1/dF^D$ (heavy green line) and $dV^*_2/dF^D$ (light blue line)
curves.
(c) $V^*_1$ (lower dark line) and $V^*_2$ (upper light red line)
for $N_{1}/N_{2} = 0.375$. 
Here no dynamical locking occurs
within this range of $F^D$. 
(d) $V^*_1$ (lower dark line) and
$V^*_2$ (upper light red line) 
for $N_1/N_2=0.292$ where dynamical locking occurs.
}
\label{fig:varyN14}
\end{figure}

We next examine the effects of varied particle ratio for a sample 
with $F^p_1=F^p_2=6.0$, $n_1=n_2$, $d/a_2=1.33$, and $n_2/N_2=0.167$.
We fix $N_2$ and vary $N_1$.
In Fig.~\ref{fig:varyN14}(a) we plot $2V_1/N$ and $2V_2/N$ versus $F^D$ for
a sample with $N_1/N_2=1/2$.
Here the velocities are normalized by $N/2$, where $N=N_1+N_2$.
In this case, channel 2 contains a larger number of particles and depins
at lower $F^D$ than channel 1.  The slope of $2V_2/N$ is also greater than
that of $2V_1/N$.
We replot the same curves in Fig.~\ref{fig:varyN14}(b) with the 
normalizations $V^*_1=V_1/N_1$ and $V^*_2=V_2/N_2$.
Under this normalization the curves have almost the same
form as the velocity force curves in samples with
an equal number of particles in each channel, and it is now possible to
distinguish the transition into a locked phase, which also appears as a
signature in the $dV^*_1/dF^D$  and $dV^*_2/dF^D$ curves shown
in Fig.~\ref{fig:varyN14}(b).
At other fillings 
such as $N_1/N_2 = 0.375$ shown in Fig.~\ref{fig:varyN14}(c),
there is no dynamical locking within this range of $F^{D}$.
For lower fillings, dynamical locking appears again as shown in 
Fig.~\ref{fig:varyN14}(d)
for $N_{1}/N_{2} = 0.292$.
We find dynamical locking only for $N_1/N_2 = 1.0$, 
$N_1/N_2=1/2$, and $N_1/N_2 < 0.333$. 

\begin{figure}
\includegraphics[width=3.5in]{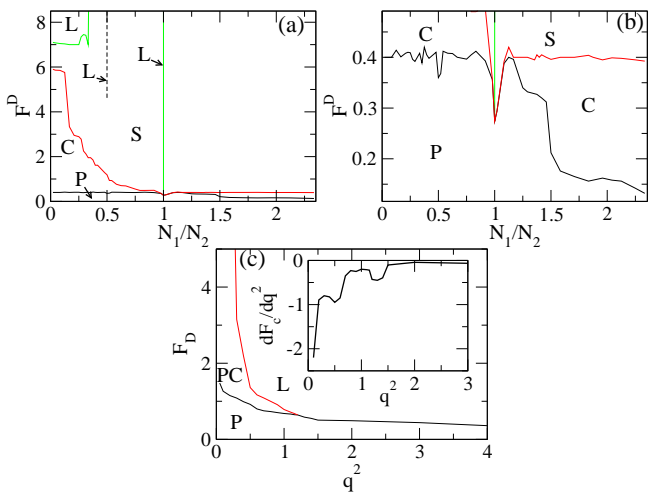}
\caption{
(a)
Dynamic phase diagram $F^D$ vs $N_1/N_2$
for the $M=2$ system in Fig.~\ref{fig:varyN14} with 
$n_1=n_2$, $n_2/N_2=0.167$,
$R_p=0.23a_0$,
$d/a_2 = 1.33$ and $F^{p}_1 = F^p_2 = 6.0$.  
P: pinned phase; C: coexistence phase; S: sliding phase; L: locked phase.
The line at $N_{1}/N_{2} = 1.0$
indicates that for this filling only the system depins elastically
directly into the L phase.
The dashed line at $N_{1}/N_{2} = 0.5$ indicates a transition from the
S phase into the L phase at this filling. 
(b) A blowup of panel (a) in the region near the depinning line
shows dips in the depinning force $F_c$
at $N_{1}/N_{2} = 0.5$ and $1.0$. 
(c) Dynamic phase diagram $F^{D}$ vs $q^2$, where $q^2$ is the dimensionless
squared colloid charge, for a sample with
$n_1=n_2$, 
$n_2/N_2=0.167$,
$N_1/N_2=1$,
$R_p=0.23a_0$,
$F^{p}_1 = F^p_2 = 6.0$, and $d/a_1 = 1.33$.
PC: the pulsed coexistence phase described in Fig.~\ref{fig:phase13}.
For strong interactions (high $q^2$) the depinning is elastic. 
For small $q^2$
the depinning threshold is high. Inset: $dF_c/dq^2$ vs $q^2$ for the same
sample
shows that at $q^2=1.15$ at the transition from plastic to elastic
depinning
there is a peak
corresponding to a 
change in the slope of the depinning curve. 
}
\label{fig:phase15}
\end{figure}

Fig.~\ref{fig:phase15}(a) shows the dynamic phase diagram
for $F^{D}$ versus $N_{1}/N_{2}$ for the system in 
Fig.~\ref{fig:varyN14}. 
The solid line at 
$N_{1}/N_{2} = 1.0$ indicates that at this ratio the
depinning is elastic and the system passes
directly from the pinned to the moving locked phase. 
For $N_{1}/N_{2} < 1.0$, channel 2 depins first since it contains a larger
number of particles, and the depinning threshold of channel 1 grows 
with decreasing $N_1/N_2$
until reaching a maximum value of $F_c=F^p_1=6.0$ at the lowest 
values of $N_1/N_2$.
The dotted line 
at $N_1/N_2=1/2$ in Fig.~\ref{fig:phase15}(a)
indicates a transition from the S to the L phase
at $F^D=4.5$ as
illustrated in Fig.~\ref{fig:varyN14}(b). 
For lower fillings $N_1/N_2 \leq 0.33$ the system enters the
locked state above $F^{D} = 7.0$. 
For $N_{1}/N_{2} > 1.0$, channel 1 depins first since it now contains
a larger number of particles than channel 2, and the depinning threshold
for channel 2 remains nearly constant as $N_1/N_2$ increases.
We note that there could be a dynamical locking at 
external drives much higher than the range of $F^D$ we consider here. 

In Fig.~\ref{fig:phase15}(b) we plot a blowup of the phase diagram 
from Fig.~\ref{fig:phase15}(a) near the depinning transition. 
Clear local minima in $F_c$ 
appear at $N_{1}/N_{2} = 1.0$
where the depinning is elastic 
and at 
$N_{1}/N_{2} = 0.5$ 
where the depinning is not elastic 
but where the dynamical locking occurs
at a much lower drive than for nearby fillings. 
The behavior at these two fillings
can be viewed as a commensuration effect, where the coupling between the layers
is enhanced at integer and certain fractional 
ratios of the filling factors. Commensuration
effects have been studied in a variety of solid-on-solid systems where there is
matching between the number of particles and the 
number of potential minima \cite{Coppersmith}.
An example of such a system is vortices in type-II superconductors with 
periodic pinning sites
where peaks in the critical depinning force occur
at integer \cite{Baret} and rational \cite{Jensen2} 
matching fields. 
In our system commensuration occurs not between the number of 
particles and the number of pinning sites
but between the number of particles in the different channels.  
The depinning force is reduced 
at the commensurate fillings 
due to the enhanced coupling between the layers.
The strongest dip in $F_c$ occurs when the depinning is elastic
at $N_{1}/N_{2} = 1.0$, while
for fillings just above and below this value
$F_c$ passes through local maxima. 
At the incommensurate fillings, there are geometrically necessary topological
defects that enter the zig-zag structure formed by the particles in the
two channels.
These defects effectively soften the structure and
reduce the coupling between the layers.
This is similar to 
what occurs in the peak effect where the depinning force is higher
for plastic depinning than for elastic depinning.
We find no apparent anomaly at $N_{1}/N_{2} = 2.0$ 
in Fig.~\ref{fig:phase15}(b);
however, we expect that for other parameter values,
more commensuration effects would be observable and that 
additional dynamical locking regimes 
would also appear at other fillings. 
Commensuration effects are also observed in two layer and three layer
systems without pinning when only one layer is driven \cite{CB}.
In these systems the commensuration effects occur at fillings
where the channels are more strongly coupled, so that the drive at which
relative slip begins to occur between the channels is much higher than for 
incommensurate fillings.
Our results indicate that a distinct
commensuration effect can occur in layered systems 
that differs from commensuration effects
observed for particles moving over fixed substrates. 
 
It is also possible to change the 
overall particle-particle interaction 
strength by varying $q^2$. 
This can be achieved for colloidal systems by altering the effective
charge on the particles or by changing the screening length. 
For vortex systems
the vortex-vortex interactions can also change significantly 
due to thermal effects near $T_{c}$. 
In Fig.~\ref{fig:phase15}(c) we plot the dynamic phase diagram 
for $F^{D}$ versus $q^2$ for the system in Fig.~\ref{fig:phase15}(a) 
with 
$N_1/N_2=1$, $F^{p}_1=F^p_2 = 6.0$ and $d/a_2 = 1.33$. 
For $q^2>1.23$ or strong particle-particle repulsion 
the system depins elastically and 
as $q^2$ increases $F_{c}$ 
gradually decreases. 
Unlike the case for constant $q^2$ but increasing density 
$1/a_1$ shown in Fig.~\ref{fig:phase13}, we find in
Fig.~\ref{fig:phase15}(c) that there
is no elastic-plastic transition at high $q^2$ 
since the zig-zag particle configuration is not affected by
increasing $q^2$.
When $q^2$ is lowered, $F_{c}$ increases and
a transition from
elastic to plastic depinning occurs at $q^2=1.15$.
For $q^2<1.15$, the system depins into the pulsed coexistence (PC) phase
described earlier in Fig~\ref{fig:phase13}.
The transition into the L phase also shifts to higher $F^D$ as
$q^2$ decreases.
The overall features of the phase diagram in Fig.~\ref{fig:phase15}(c)
resemble those of the dynamic phase diagrams constructed for 2D 
vortex systems when the vortex-vortex interaction strength is 
varied \cite{Olson,N}.  In the vortex case,
$F_{c}$ increases as the vortex-vortex interactions become weaker while
the dynamical ordering transition shifts to higher values 
of the driving force. 
In the inset of Fig.~\ref{fig:phase15}(c) we plot 
$dF_{c}/dq^2$ versus $q^2$, showing more clearly the change in the
slope of $F_{c}$ at 
the transition from plastic to elastic depinning at
$F^D=1.15$. 

\begin{figure}
\includegraphics[width=3.5in]{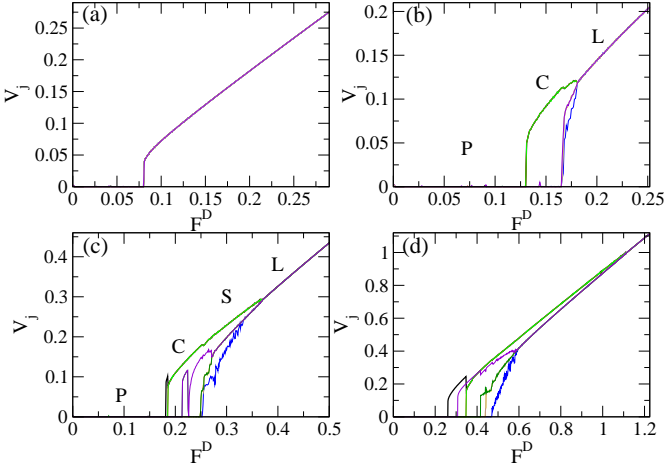}
\caption{
$V_j$ vs $F^D$ for an $M=8$ system with
$N_j=N_1$, $n_j=n_1$, $n_1/N_1=0.17$, 
$R_p=0.23a_0$, 
$a_1=1.5a_0$, 
and $d/a_1=1.33$.
(a) 
At $F^{p}_j=F^p_1 = 3.0$, 
the depinning is elastic and all channels
depin simultaneously into the locked state. 
(b) At $F^{p}_j=F^p_1 = 5.0$, plastic depinning occurs
and groups of channels 
lock together with each group depinning at a different value of $F^D$.
At high drives all the channels lock.
P: pinned phase; C: coexistence phase, defined as beginning when the first
channel depins and ending when the final channel depins; L: locked phase.
(c) At $F^{p}_j=F^p_1 = 7.0$, more channels depin individually. 
A window of the sliding (S) phase appears where all channels are moving
but groups of channels are separately locked.
(d) At $F^{p}_j=F^p_1 = 11.0$, the same four dynamical phases appear but are
shifted to higher values of $F^{D}$. 
}
\label{fig:vel16}
\end{figure}

\section{More Than Two Channels}
We next consider the case of $M>2$ coupled driven channels. 
Figure~\ref{fig:vel16} 
shows the velocity-force curves $V_1$ through $V_8$ versus $F^D$
for an $M=8$ channel system
with $N_j=N_1$, $n_j=n_1$, $d/a_1 = 1.33$,  and 
$n_{1}/N_1 = 0.17$. 
Note that the boundary transverse to the channels is 
open and that there is therefore no
significant direct interaction between the particles in channel 1 and 
those in channel 8.
In Fig.~\ref{fig:vel16}(a) at $F^{p}_j=F^p_1 = 3.0$, 
the depinning is elastic and the system immediately enters the
locked phase upon depinning. All the channels
depin simultaneously and the velocity curves overlap completely.
In Fig.~\ref{fig:vel16}(b) at $F^{p}_j=F^p_1 = 5.0$ 
the depinning is plastic.  Channels 1, 2, and 3 depin first and remain locked
with each other, followed by the depinning of a group of four additional locked
channels and then by the depinning of the final channel.
At $F^{D} = 0.182$ all the channels become dynamically locked together.
For $F^{p}_j=F^p_1 = 7.0$ in Fig.~\ref{fig:vel16}(c), 
the system exhibits a hierarchy of 
dynamical locking transitions beginning at $F^D=0.186$ when the first three
channels to depin become locked together.
This is followed by a second dynamic locking of four different channels
at $F^{D} = 0.25$. 
Channel 8 dynamically locks with these four channels
near $F^{D} = 0.335$, and a final locking of all the channels with
each other occurs
at $F^{D} = 0.37$. 
As $F^p$ increases the number of distinct dynamical locking transitions
increases, as illustrated in Fig.~\ref{fig:vel16}(d) for $F^p_j=F^p_1=11.0$,
while the final transition at which all of the channels lock together
shifts to higher values of $F^D$.
Figure~\ref{fig:vel16}(d) also shows that it is possible for
individual channels to exhibit negative differential conductivity at 
the onset of channel locking; however, there is generally no NDC
in the summed velocity $V=\sum_j^{M}V_j$.

\begin{figure}
\includegraphics[width=3.5in]{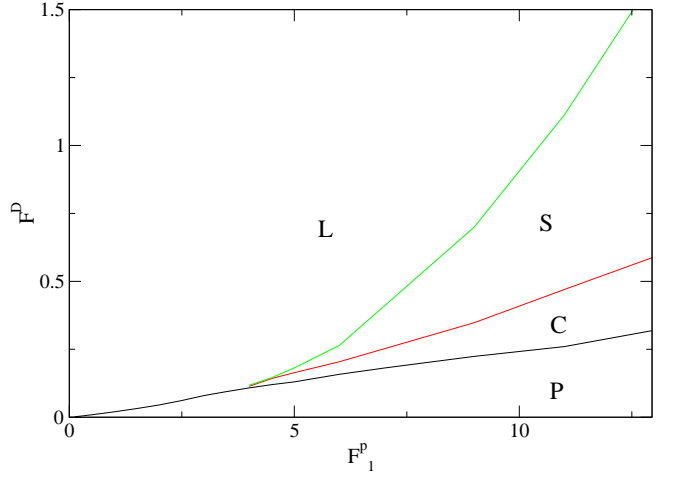}
\caption{
Dynamic phase diagram $F^D$ vs $F^p_1$ for the $M=8$ system in 
Fig.~\ref{fig:vel16} with $N_j=N_1$, $n_j=n_1$, $F^p_j=F^p_1$, $n_1/N_1=0.17$,
$R_p=0.23a_0$, $a_1=1.5a_0$, and $d/a_1=1.33$.
P: pinned phase; C: coexistence phase; S: sliding phase; L: locked phase.
}
\label{fig:phase17}
\end{figure}

\begin{figure}
\includegraphics[width=3.5in]{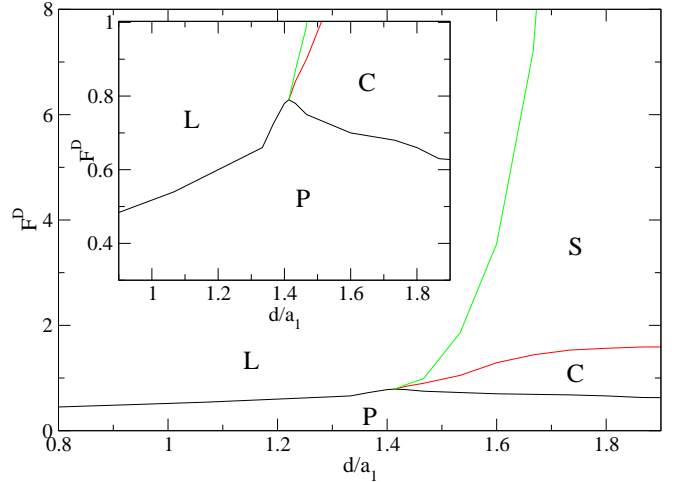}
\caption{
Dynamic phase diagram $F^D$ vs $d/a_1$ for the 
$M=8$ system in Fig.~\ref{fig:phase17}
with $F^{p}_j=F^p_1 = 3.0$, 
$N_j=N_1$, $n_j=n_1$, $n_1/N_1=0.17$,
$R_p=0.23a_0$, and fixed $a_1=1.5a_0$.
P: pinned phase; C: coexistence phase; S: sliding phase; L: locked phase. 
Inset: Blow up of
the region near the transition from elastic to plastic depinning
shows that the depinning force $F_c$ peaks at the transition.
}
\label{fig:phase18}
\end{figure}

Figure~\ref{fig:phase17} shows
the dynamic phase diagram $F^{D}$ versus $F^{p}_1$ for the
$M=8$ system in Fig.~\ref{fig:vel16} with $F^p_j=F^p_1$. 
The broad features of the phase diagram are the same as those found
for the $M=2$ case in Fig.~\ref{fig:phase12}; however, there are multiple
depinning transitions within the C phase and multiple locking 
transitions within the S phase that are not marked in Fig.~\ref{fig:phase17}.
In Fig.~\ref{fig:phase18} we plot the dynamic phase diagram
$F^D$ versus $d/a_1$ for the same $M=8$ system with
$F^p_j=F^p_1=3.0$ and fixed $a_1=1.5a_0$.
At low $d/a_1$ the system depins elastically, and a transition to plastic
depinning occurs near $d/a_1=1.4$.
For $d/a_1>1.4$, the transition between the S and L phases shifts to higher
$F^D$ with increasing $d/a_1$, while the general shape of the phase diagram
resembles that of the $M=2$ system shown in Fig.~\ref{fig:phase3}.
The C and S phases again contain multiple depinning and dynamical locking
transitions, respectively.
We observe a larger amount of hysteresis in $M>2$ systems than in the
$M=2$ system.
In the inset of Fig.~\ref{fig:phase18} we plot a blowup of
the $F^D$ versus $d/a_1$ phase diagram near the depinning transition to
show that $F_c$ passes through a peak at the transition from elastic to
plastic depinning, similar to the peak found for the $M=2$ system 
in the inset of Fig.~\ref{fig:phase3}(a).

\begin{figure}
\includegraphics[width=3.5in]{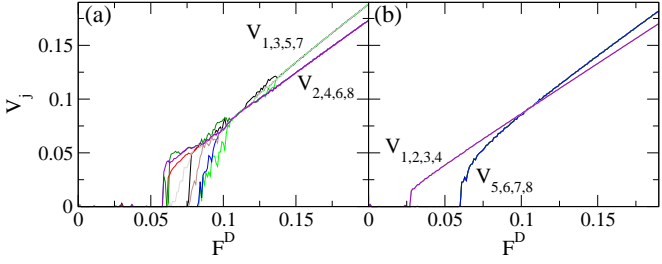}
\caption{
$V_j^*$ vs $F^D$ for an $M=8$ system with $N_j \neq N_1$,
$n_j=n_1$, $F^p_j=F^p_1=2.0$, 
$n_j/N^{\rm high}=0.167$, $R_p=0.23a_0$, and $d=1.7a_0$.
Each channel has one of two possible densities: 
$d/a_j=1.33$ for $N_j=N^{\rm high}$ or
$d/a_j=1.2267$ for $N_j=N^{\rm low}$.
All of the velocities are normalized by $V_j^*=V_j/N^{\rm high}$.
(a) 
A sample with alternating density 
$N_1=N_3=N_5=N_7=N^{\rm high}$ and
$N_2=N_4=N_6=N_8=N^{\rm low}$.
All eight channels depin at different values of $F^{D}$
and undergo a series of locking and unlocking transitions until
reaching a state at higher drives where all the low density channels are
locked together and all the high density channels are locked together.
(b) A sample with the same parameters but with segregated density 
$N_1=N_2=N_3=N_4=N^{\rm low}$ and
$N_5=N_6=N_7=N_8=N^{\rm high}$ shows a two
step depinning process.  
}
\label{fig:iv19}
\end{figure}

For systems in which the number of particles per channel is allowed to
vary, $N_j \neq N_1$, the dynamic behavior becomes more complex; however,
it is possible to identify several general features.
In Fig.~\ref{fig:iv19} we plot $V_{j}^*$ versus $F^{D}$ for
an $M=8$ channel with $F^p_j=F^p_1=2.0$ and $d=1.7a_0$ where each channel
contains either $N^{\rm high}$ particles with $d/a_j=1.33$ or $N^{\rm low}$ 
particles with 
$d/a_j=1.2267$.
The velocities are normalized by $V_j^*=V_j/N^{\rm high}$, and the number of
pins in each channel is $n_j/N^{\rm high}=1.67$.
In samples where all channels have equal numbers of particles, 
$N_j=N_1=N^{\rm high}$ or $N_j=N_1=N^{\rm low}$, the depinning is elastic for
this set of parameters.
The depinning becomes plastic when the number of particles differs from
channel to channel.
Figure~\ref{fig:iv19}(a) illustrates an alternating density system
with $N_1=N_3=N_5=N_7=N^{\rm high}$ and $N_2=N_4=N_6=N_8=N^{\rm low}$.
Here every channel has a unique depinning threshold.
There are multiple locking and unlocking transitions as channels become
locked with a nearest neighbor channel at lower drives only to unlock at
higher drives.
The locking at lower drives occurs when the pinning is still able
to create density inhomogeneities within individual channels.  Particles
form higher density regions in the areas behind the more effective pins where
pileups occurred below the depinning transition.  These higher density regions
create an additional energy barrier for relative slip between the particles
in neighboring channels, favoring the coupling of adjacent channels.  
This process creates a window $0.105 < F^D < 0.115$ where all of the channels
couple together, $V_j=V_1$.
As $F^D$ increases, the effectiveness of the pinning is reduced and the density
within each channel becomes more homogeneous.  At the same time, the overall
effectiveness of the pinning in channels with $N^{\rm high}$ is reduced 
relative to that of the pinning in channels with $N^{\rm low}$ due to the 
stronger particle-particle interactions in the higher density channels.
The velocity $V^{\rm high}$ in channels with $N^{\rm high}$ is therefore higher
than 
the velocity $V^{\rm low}$ in channels with $N^{\rm low}$,
$V^{\rm high}>V^{\rm low}$, which favors decoupling of adjacent channels in
the alternating density system.
For high enough drives $F^D>0.14$, the 
system enters a doubly locked phase where all channels containing the same
number of particles are locked together, $V_1=V_3=V_5=V_7$ and
$V_2=V_4=V_6=V_8$, but $V_1>V_2$ by a small amount so that the two sets of 
locked channels slip with respect to each other.

There are many other possible ways to arrange the $M=8$ channels such that
half of the channels contain $N^{\rm high}$ particles and half of the
channels contain $N^{\rm low}$ particles.
For example, Fig.~\ref{fig:iv19}(b) shows 
$V_j^*$ versus $F^D$ in a density segregated system
with $N_1=N_2=N_3=N_4=N^{\rm low}$ and $N_5=N_6=N_7=N_8=N^{\rm high}$.
Here each group of channels with equal density acts as a unit and depins
elastically into a locked moving state; however, the two groups of 
channels do not lock with each other.  The depinning occurs in two
steps with a lower depinning threshold for the set of channels containing
$N^{\rm low}$ particles.
This behavior resembles that of an $N=2$ sample with $N_1/N_2$ slightly
below 1, as in Fig.~\ref{fig:phase15}(b), which depins into the coexistence
phase and then enters the sliding phase at higher $F^D$.
In Fig.~\ref{fig:iv19}(b) we find that $V_1^*>V_5^*$ just above the second
depinning transition but $V_1^*<V_5^*$ at high $F^D$, with a crossing
of the velocity curves occurring near $F^D=0.105$.

\begin{figure}
\includegraphics[width=3.5in]{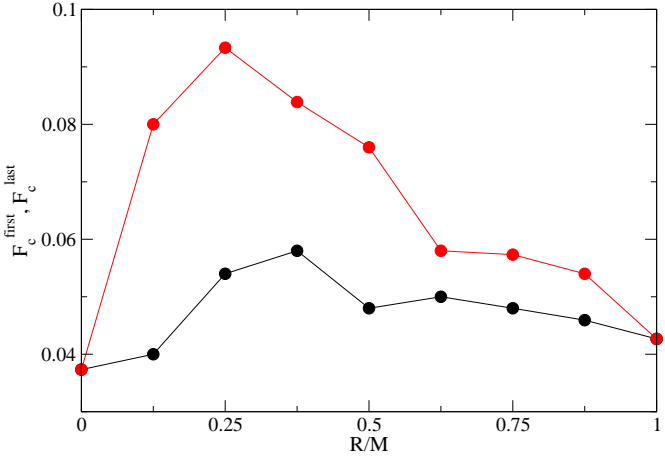}
\caption{
The depinning thresholds $F_c^{\rm first}$ (lower curve) and 
$F_c^{\rm last}$ (upper curve) vs 
$R/M$
for the $M=8$ sample from Fig.~\ref{fig:iv19}(a)
with $n_j=n_1$, $F^p_j=F^p_1=2.0$, $n_j/N^{\rm high}=0.167$,
$R_p=0.23a_0$, $d=1.7a_0$, $d/a_j=1.133$ for $N_j=N^{\rm low}$,
and $d/a_j=1.216$ for $N_j=N^{\rm high}$.
There are $R$ evenly spaced channels with $N_j=N^{\rm high}$ and the
remaining $M-R$ channels have $N_j=N^{\rm low}$.
When the system is commensurate 
with $N_j=N_1$ at $R/M=0$ or $R/M=1$, the depinning is elastic.
For other values of $R/M$ the depinning is plastic, and the maximum depinning
thresholds appear when geometrically necessary dislocations form in
the system.
}
\label{fig:depin20}
\end{figure}

In samples with an equal number of particles per channel, particles in
neighboring channels can organize into a two-dimensionally ordered state in
order to minimize the particle-particle interaction energy.  This ordered
configuration can be perturbed by changing the number of particles in one or
more channels, which causes defects to appear.
To study this, we consider an $M=8$ system in which all channels 
initially contain
$N_j=N^{\rm low}$ particles with $d/a_j=1.133$.
We select $R$ channels evenly spaced across the system
with $0 \leq R \leq M$
and increase the number of particles in the selected channels to
$N_j=N^{\rm high}$ with $d/a_j=1.216$.  
As $R$ varies, the system passes from a state in
which all channels have $N_j=N^{\rm low}$ at $R/M=0$ 
to one in which all channels have
$N_j=N^{\rm high}$ at $R/M=1$.  
Fig.~\ref{fig:depin20} 
shows $F_c^{\rm first}$, the value of $F^D$ at which the first channel depins,
and $F_c^{\rm last}$, the value of $F^D$ at which the final channel depins,
for different values of $R/M$.
The geometrically necessary dislocations that form for
intermediate values of $R/M$
are aligned with the direction of the drive.
At $R/M=0$ and 1, dislocations are absent and
the depinning transition is elastic, but for
other values of $R/M$ when dislocations are present, the depinning
is plastic and the depinning thresholds are shifted to higher
values of $F^D$.
This can also be seen in Fig.~\ref{fig:iv19}: the alternating filling
sample in Fig.~\ref{fig:iv19}(a) contains numerous 
topological defects and has much
higher depinning thresholds than the segregated filling sample in
Fig.~\ref{fig:iv19}(b), where there are only dislocations along the
interface between the two fillings.
This result again indicates that in our model
the higher depinning forces are 
associated with plastic depinning 
where the channels can slide past
each other due to the presence of topological defects created either by
geometrical necessity for channels with unequal particle numbers or by
the quenched disorder.
The results in Fig.~\ref{fig:iv19} and Fig.~\ref{fig:depin20} 
are similar to previous work on mixtures of particles 
in the presence of a weak disordered substrate, which showed
that the depinning force is minimized when only one particle species is
present and there are no topological defects, 
but that for intermediate mixtures of particle species,
topological defects appear and increase the depinning force \cite{M}.      

\begin{figure}
\includegraphics[width=3.5in]{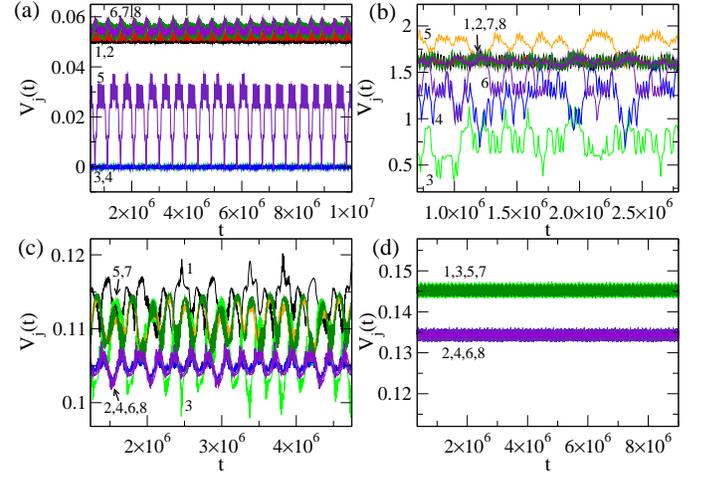}
\caption{
The time trace of the velocities of the individual channels,
$V_j(t)$, for the $M=8$ sample from 
Fig.~\ref{fig:iv19}(a) 
with $N_1=N_3=N_5=N_7=N^{\rm high}$ and $N_2=N_4=N_6=N_8=N^{\rm low}$.
(a) At $F^{D} = 0.08$ the system is in the coexistence phase.
Channels 3 and 4 are pinned,
channel 5 flows intermittently,
channels 1 and 2 are locked at a low average velocity, and
channels 6, 7, and 8 are locked at a higher average velocity.
(b) At $F^{D} = 0.09$ there is considerably more intermittency. 
Here channels 1, 2, 7 and 8 are locked continuously, while channels
3, 4, 5, and 6 show intermittent locking with the other channels.
(c) At $F^{D} = 0.125$ there is less intermittency and a larger
number of the channels are locked with other channels.
(d) At $F^{D} = 0.16$, channels 1, 3, 5, and 7 are locked and due to their
higher density move with a higher velocity than the remaining 
channels 2, 4, 6, and 8, which are locked together.
}
\label{fig:lock21}
\end{figure}

The locking-unlocking transition that occurs for the alternating density
sample in
Fig.~\ref{fig:iv19}(a)
can be better characterized by analyzing the time series $V_j(t)$ of the 
velocities of the individual channels
at different values of $F^{D}$.
In Fig.~\ref{fig:lock21}(a), where we plot $V_j(t)$ at $F^{D} = 0.08$, 
the system is in the coexistence regime and
channels 3 and 4 are pinned.
Channel 5 switches between a temporarily pinned state and a sliding
state, with occasional jumps to higher velocity flow.
Channels 1 and 2 are locked together and have an average
velocity of $V_1=V_2=0.052$, while channels 6, 7, and 8 are locked together
at a slightly higher average velocity of $V_6=V_7=V_8=0.055$
and show a pronounced modulation at a frequency which matches the
frequency at which $V_5$ drops to zero.
The point at which $V_5$ reaches zero corresponds to the point at
which $V_6$, $V_7$, and $V_8$ reach their maximum values.
Numerous periodic oscillations appear in the velocities, such as the
washboard frequency of the particles moving over the pinning sites.
Additionally, each channel has a periodicity determined by the value of
$N_j$, and since this frequency differs from channel to channel,
the channels sliding past one another experience
a dynamic periodic potential caused by the periodicity of the particles.
This slipping process is not completely periodic but shows some
changes over time. 
For lower drives, more channels are pinned and the 
velocity signals become increasingly periodic. 
We note that for the $M=2$ case 
illustrated in Fig.~\ref{fig:sofw8} the velocity time traces were dominated
by the fundamental frequencies which were washboard signals.

Figure~\ref{fig:lock21}(b) shows $V_j(t)$ in the same sample for $F^{D} = 0.09$. 
Here there is considerable intermittency and several channels lock and
unlock.
Channels 1, 2, 7, and 8 are locked and move at
$V_1=V_2=V_7=V_8=0.0625$.
Channel 3 has the lowest average velocity and intermittently locks with
channel 4, which has the second lowest average velocity.
Channel 4 locks intermittently with both channels 3 and 6, as well as
much more infrequently locking with the group of channels 1, 2, 7, and 8.
Channel 6 also locks intermittently with the channel group 1, 2, 7, and 8 as
well as with channel 4.
Channel 5 has the highest average velocity and intermittently locks to the
lower velocity value of the group of channels 1, 2, 7, and 8.
In Fig.~\ref{fig:lock21}(c) at $F^D=0.125$, the amount of intermittency is
reduced.  Channels 2, 4, 6, and 8 are locked together and have a lower
average velocity, while channels 5 and 7 lock together and have a higher but
strongly fluctuating velocity.  Channel 1, which moves independently, reaches
the highest velocity values, while channel 3, which also moves independently,
reaches the lowest velocity values.
At $F^{D} = 0.16$, shown in Fig.~\ref{fig:lock21}(d), 
the high density channels 1, 3, 5, and 7 are locked together and move at
a higher average velocity while the low density channels 2, 4, 6, and 8 also
lock with each other and move at a lower average velocity.

The multiple channel system exhibits many of the features 
predicted in mean field studies of phase slip systems \cite{Saunders},
such as the coexistence of sliding and pinned phases.
There are, however, many features of the multiple channel system
that are not captured by the mean field results,
such as the multiple coupling and decoupling transitions between pairs
of channels or groups of channels or the peak effect in the depinning
threshold at the transition from elastic to plastic depinning.
Additionally, for both the two channel and multiple channel
systems, the case where different amounts of disorder exist 
in each channel has not been 
studied with mean field models. 
We note that that there are many 
more parameters that can be explored for the multiple
channel systems, such as having different densities or 
pinning strengths for each channel;
however, the results for the dynamic phase diagram for the multiple channel
system suggest that many of the generic features
observed for the two level system will persist in the
multiple channel system. We will explore this in more detail elsewhere.

\section{Summary} 

We have examined systems of two or multiple one-dimensional channels 
of coupled particles that interact repulsively within each channel
and between the channels. 
The particles are uniformly driven with an external drive
and in the presence of quenched disorder show a series of dynamic
phases including a pinned phase, 
a coexistence between pinned and sliding phases,
a sliding phase where the channels move at different average velocities
and slip past one another, and
a dynamically locked phase where the particle positions in the channels become
locked and the channels move at the same velocity. The transition between
these different phases
can be observed as clear features 
in the velocity force curve characteristics.
These features include a sudden drop in the velocity or the onset of 
negative differential conductivity in one of more of the channels at the
dynamically induced locking transition.  
For weak pinning or strong channel coupling, the depinning occurs
elastically without any slipping between channels. 
When the system parameters are varied, such as by decreasing the coupling 
between channels, we find a transition to plastic depinning, with the
initial flow above depinning in the coexistence regime where the
channels depin separately. 
At the transition between elastic and plastic depinning,
a peak in the depinning force appears which resembles
the peak effect phenomenon found in more complicated models 
with transitions between elastic and plastic depinning. 
We have also examined   
channels containing unequal numbers of pinning sites or particles and find
that even for a channel containing no pinning, there can be a finite
depinning threshold 
due to the coupling
of the particles in the pin-free channel with the particles in
a channel containing pinning. 
For unequal numbers of particles in the channels we observe
commensurability effects where the depinning threshold drops when the
depinning becomes elastic for particle number ratios 
at which the coupling between particles in neighboring channels is
enhanced, such as at
1:1 or 2:1.
At the incommensurate
fillings the channels always depin plastically.
For multiple channels we find a hierarchy of dynamically induced 
coupling phases in which different groups of channels couple, while
at higher drive all of the channels become coupled. 
Despite the apparent simplicity of our model, 
we find that even the two channel system exhibits
many of the prominent features of depinning in more complex
models such as the existence of multiple dynamical 
phases as well as elastic to plastic depinning
transitions of the type that have been studied for
more complicated systems including 3D layered systems. 

The specific model studied in this paper
could be realized using colloidal particles in coupled
1D channels or in superconductors with a one-dimensional  
corrugation combined
with random pinning where the vortices move along the easy flow direction
of the corrugation. Other possible realizations
include coupled channels of 1D 
wires where Wigner crystallization may occur.  
 
\section{Acknowledgments}
This work was 
carried out under the auspices of the NNSA of the U.S. DoE at LANL
under Contract. No. DE-AC52-06NA25396.

\end{document}